\pdfoutput=1
\documentclass[useAMS,usenatbib,referee]{mn2e}
\usepackage {graphicx} 
\def\apj{ApJ\,}
\def\apjl{ApJ\,}
\def\aap{A\&A\,}
\def\mnras{MNRAS\,}
\def\pasj{PASJ\,}
\def\aj{AJ}
\def\aaps{A\&AS}
\def\pasp{PASP}

\title     {Scaling for the  intensity of radiation in 
            spherical and aspherical  planetary nebulae}
\author    [L. Zaninetti] 
           {L. Zaninetti  \\ 
            Dipartimento di Fisica Generale, \\
           Via Pietro Giuria 1               \\
           10125 Torino, Italy               \\
  email:  zaninetti@ph.unito.it
}
\date      {~~~~~~~~~~~~~~~~~~~~~~~~~~}
\date{to be inserted}

\pagerange{\pageref{firstpage}--\pageref{lastpage}}
\pubyear{2008}
\everymath     {\rm}
\everydisplay  {\rm}
\begin{document}

\maketitle

\label{firstpage}

\begin{abstract}
The image of planetary  nebulae  is made by   three different
physical processes. 
The first process is the expansion of the shell that can be 
modeled by the canonical  laws of motion in the spherical 
case and   by the momentum conservation 
when 
gradients of density
are 
present
in the interstellar medium.
The second process is the diffusion 
of  particles that radiate from the advancing layer.
The 3D diffusion from a sphere as well as the 1D diffusion 
with drift are analyzed.
The third process is the composition of the image
through an integral operation along  the line
of sight.
The developed framework is  applied to A39 , to the Ring 
nebula and to the etched hourglass nebula  MyCn 18.

\end{abstract}


\section{Introduction}

The planetary nebula , in the following PN , rarely 
presents   a circular shape generally thought
to be the projection of a sphere on the sky.
In order to explain the properties of  PN,
\cite{Kwok1978} proposed the interacting 
stellar wind (ISW) theory.
Later on \cite{Sabbadin1984}  
proposed the two wind 
model and the two phase model.
More often various types of shapes 
such as elliptical , bipolar or cigar are present,
see~\cite{Balick1987,Schwarz1992,Manchado1996,Guerrero2004,Soker2002a,
Soker2002b}.
The bipolar PNs , for example , are explained by the interaction
of the winds which   originate  from the central star ,
see~\cite{Icke1988, Frank1995, Langer1999, Gonzales2004a}.
Another class of models explains some basic structures 
in PNs through hydrodynamical models, 
see~\cite{Kahn1985,Mellema1991} or 
through  self-organized magnetohydrodynamic (MHD) plasma 
configurations with radial
flow, 
see~\cite{Tsui2008}. 

An attempt 
to make a catalog of line profiles 
 using various shapes  observed in real PNs
was  done by \cite{Morisset2008}.
This ONLINE atlas , available
at 
\newline
 http://132.248.1.102/Atlas$\_$profiles/img/,
is  composed
of  26 photo-ionization models corresponding to 5 geometries,
3 angular density laws and 2 cavity sizes,
 four velocity fields 
for  
a total of  104 PNs,
each of which can be observed from 3 different
directions.

\cite{Matsumoto2006} suggest  that
a planetary nebula is formed and evolves by the interaction
of a fast wind from a central star with 
a slow wind from its progenitor , an  
Asymptotic Giant  Branch (AGB) star.
It seems therefore reasonable to assume that the PN evolves 
in a previously ejected medium 
( AGB) phase in which density is considerably 
higher  than the interstellar medium (ISM) .
We can , for example ,consider a  PN resulting from
a 5 $M_{\sun}$ Main Sequence (MS) star .
The central core will be a White Dwarf (WD) less than
1 $M_{\sun}$ and the ionized nebula is generally
less than  1 $M_{\sun}$ . 
We therefore have $\approx$ 3 $M_{\sun}$ of gas around the
PN which   come  from the AGB.
The number density that  characterizes the PN is 
\begin{equation}
n \approx  \frac {9.66 M_{1,\sun} }{R_{pc}^3}
\frac{particles}{cm^3}
\quad ,
\label{concagb}
\end{equation} 
where $ M_{1,\sun}$ is the number of solar masses 
in the volume occupied by the nebula
and $R_{pc}$ the radius of the nebula in pc.

By inserting 
$M_{\sun}$=0.605,
see  for example  Figure 2 in  \cite{Steffen2004} , 
 and  $R_{pc}$=1
in the previous formula 
we obtain
$n \approx 6.28    \frac{particles}{cm^3} $ .
This can be considered an averaged value 
and it  should be noted  that the various hydrodynamical models 
give densities ,$\rho$ ,  that  scale 
with the distance from the center $R$
as
$R^{-\alpha}$ ,
with $2.5~<\alpha<3.5$ ,
see
      \cite{Villaver2002}, 
      \cite{Steffen2004}, 
      \cite{Steffen2005a},
      \cite{Steffen2005b},
      \cite{Steffen2007},
and  \cite{Steffen2008}. 
\label{secalfa}

The already cited  models concerning the PNs 
leave a series of questions
unanswered or partially answered:
\begin {itemize}
\item Which are the laws of motion that regulate
      the expansion of PN~?
\item  Is it possible to build up a diffusive model 
       in the thick advancing layer~?
\item  Is it possible to deduce some analytical formulas 
       for the intensity profiles~?
\end{itemize}

In order  to answer these questions  Section~\ref{sec_pn}
describes three observed morphologies of PNs,
Section~\ref{sec_motion}  analyzes three different
laws of motion that model the spherical and aspherical expansion,
Section~\ref{sec_diffusion} reviews old and new formulae
on diffusion 
and Section~\ref{sec_image} contains detailed information
on how to build an image of a PN.

\section{Three morphological types of PNs} 
\label{sec_pn}
This section presents the astronomical data of a nearly
spherical PN  known as  A39, a weakly asymmetric shell , the Ring nebula ,
and a bipolar PN which is the etched hourglass nebula  MyCn 18 .

\subsection{A circular spherical PN}

\begin{figure*}
\begin{center}
\includegraphics[width=8cm]{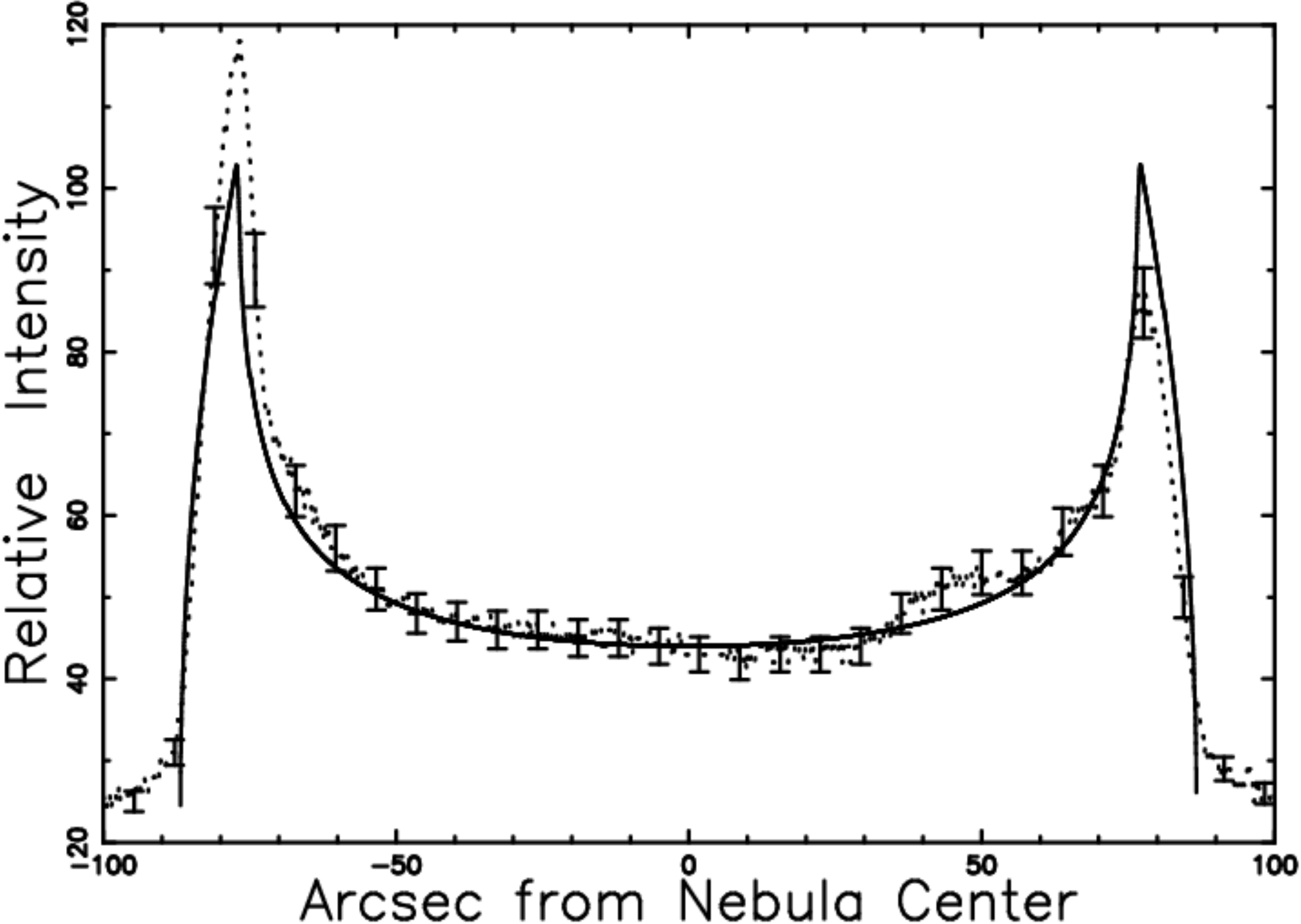}
\end {center}
\caption
{
 Cut of the relative intensity of A39 
 crossing the center    
 in the east-west direction  (dotted line with some error bar) 
 and  the rim model          (full  line) 
 fully described in ~Jacoby et al. (2001) ,
 $\chi^2$ = 0.505 when 5819 point are considered.
}
\label{cuta39}
    \end{figure*}

The PN A39 is extremely round and therefore 
can be considered  an example of spherical symmetry,
see for example Figure~1 in \cite{Jacoby2001} .
In A39  the radius  of the shell , $R_{shell}$  is 
\begin{equation}
 R_{shell} = 2.42 \times 10^{18} \Theta_{77} D_{21} ~cm
=0.78~pc
\quad ,
\end{equation}
where $\Theta_{77}$ is the angular radius in units  of
$77^{\prime\prime}$ and  $D_{21}$ the distance in units 
of  2.1~kpc ,  see \cite{Jacoby2001}~.
The expansion velocity has a range $[32 \leftrightarrow  37~\frac {km}{s} ]$ 
according to  \cite{Hippelein1990} and the age  of  the 
free  expansion is 23000 yr, see \cite{Jacoby2001}.
The angular thickness of the shell is
\begin{equation}
\delta\,r_{shell} = 3.17\; 10^{17} \Theta_{10} D_{21} cm =0.103 ~pc
\quad  ,
\end{equation}
where $\Theta_{10}$ is the thickness  in units  of
$10.1^{\prime\prime}$ and 
the height above the galactic plane is 1.42 $kpc$ ,
see \cite{Jacoby2001}.
The radial distribution of the intensity in $[OIII]$ image 
 of  A39
after  subtracting  the  contribution
of the central star
is well described
by a  spherical
shell with a $10^{\prime\prime}$  rim thickness,
see Figure \ref{cuta39}  and  \cite{Jacoby2001}.

The caption of Figure~\ref{cuta39} 
also reports  the $\chi^2$ of the fit
computed according to formula~(\ref{chi2}). 
In presence of real data a merit function ,$\chi^2$ , 
is  introduced as 
\begin{equation}
\chi^2 =\frac{1}{N}  \sum_{i=1} ^N \bigl [ \frac { y_i - y_{i,obs}} {\sigma_i} \bigr ]^2
\quad ,
\label{chi2}
\end {equation}
where $N$ is the number of the data , 
$y_{i}$ the theoretical ith  point  ,
$y_{i,obs}$ the ith observed point and
$\sigma_i$ 
the error for the ith observed point
here computed as $\frac{y_{i}}{10}$.

\subsection{The asymmetric PN}

The Ring nebula ,
also known as M57 or NGC6720 , 
presents an elliptical  shape  characterized by a semi-major axis  of 
$42 ^{\prime\prime}$,  
a semi-minor  axis  of 
$29.4 ^{\prime\prime}$
and ellipticity of  0.7,
see Table I in \cite {Hiriart2004}.
The distance of the Ring nebula is not very well known ;
according to \cite {Harris1997} the distance is  
705 $pc$ .
In physical units the two radii are 
\begin{eqnarray}
 R_{shell,minor} = 0.1  \Theta_{29.4} D_{705} ~pc \quad ~~~ semi-minor ~radius
\nonumber \\
 R_{shell,major } = 0.14  \Theta_{42} D_{705} ~pc \quad ~~~ semi-major ~radius
\quad  , 
\end{eqnarray}
where $\Theta_{29.4}$ is the angular minor radius in units  of
$29.4^{\prime\prime}$,
$\Theta_{42}$ is the angular major  radius in units  of
$42^{\prime\prime}$  and  
$D_{705}$ the distance in units 
of  705~$pc$.
The radial velocity structure 
in the Ring Nebula  
 was
derived from observations of the 
$H_2$ (molecular Hydrogen) ~ v = 1- 0 S(1) emission
line at 2.122 $\mu m$ 
obtained by using a cooled Fabry-
Perot etalon and a near-infrared imaging detector
, see  \cite{Hiriart2004} .
The velocity structure of the Ring Nebula
covers  the range 
$[-30.3 \leftrightarrow 48.8~\frac {km}{s} ]$~.

\subsection{The case of   MyCn 18  }

MyCn 18 is a PN at a distance of  2.4 $kpc$ 
and  clearly shows an hourglass-shaped nebula,
see  \cite{Corradi1993,Sahai1999}.
On  referring  to Table~1 in~\cite{Dayal2000} we can  fix 
the equatorial radius in  $2.80 \times 10^{16}~cm$ ,
or   $0.09~pc$   ,
and the radius at $60^{\circ}$ from the equatorial plane 
$3.16 \times 10^{17}~cm$
or $0.102~pc$~.
The determination of the  observed field of velocity 
of MyCn 18 varies  from an overall value
of 10 $\frac{km}{s}$ as  suggested   by the expansion
of $[OIII]$ , see \cite{Sahai1999} ,
to a theoretical model by  \cite{Dayal2000} in which
the velocity is  9.6 $\frac{km}{s}$ when
the latitude is 0 $^{\circ }  $ (equatorial plane)
to
40.9  $\frac{km}{s}$ when
the latitude is 60 $^{\circ }  $.

\section{Law of motion}
\label{sec_motion} 
This Section presents two solutions for the law of motion 
that describe  asymmetric expansion.
The momentum conservation is then applied in  cases
where  the density of the interstellar medium is not
constant but regulated by  exponential behavior.

\subsection{Spherical Symmetry - Sedov solution}

The momentum conservation is applied  to a conical  section 
of radius  $R$ with  a solid   angle   $\Delta\;\Omega$, 
in polar coordinates, see   ~\cite{Dalgarno1987} 
\begin {equation}
\frac {d}{dt} (\Delta M  R)  = \Delta  F 
\,,
\end {equation}
where
\begin {equation}
\Delta M  =  \int_0^R \rho   (R,\theta,\phi) dV 
\,,
\end {equation}
is the mass  of  swept--up  interstellar   medium 
in the  solid angle  $\Delta\;\Omega$,
$\rho$  the density  of the medium ,
$P$ the interior pressure 
  and the   driving force:
\begin {equation}
\Delta \; F =  PR^2 \Delta \Omega  
\,.
\end  {equation}
After some algebra the Sedov  solution    is 
obtained, see~\cite{Sedov1959,Dalgarno1987} 
\begin{equation}
R(t)=
\left ({\frac {25}{4}}\,{\frac {{\it E}\,{t}^{2}}{\pi \,\rho}} \right )^{1/5}
\quad , 
\label{sedov}
\end{equation}
where $E$ is the  energy injected in the process
and  $t$ the  time.

Another slightly different solution is 
formula~(7.56) in  \cite{Dyson1997} 
\begin{equation}
R(t)=
\left ({\frac {25}{3}}\,{\frac {{\it E}\,{t}^{2}}{\pi \,\rho}} \right )^{1/5}
\quad , 
\label{sedovdue}
\end{equation}
where the difference is due to the adopted approximations.

Our astrophysical  units are: time ($t_4$), which
is expressed  in $10^4$ \mbox{yr} units;
$E_{42}$, the  energy in  $10^{42}$ \mbox{erg}; and    $n_0$  the
number density  expressed  in particles~$\mathrm{cm}^{-3}$~
(density~$\rho=n_0$m, where m=1.4$m_{\mathrm {H}}$).
With these   units equation~(\ref{sedov}) becomes
\begin{equation}
R(t) \approx  0.198
\left ({\frac {{\it E_{42}}\,{t_4}^{2}}{n_0}} \right)^{1/5}~pc  
\quad . 
\label{sedovastro}
\end{equation}
The expansion velocity is
\begin{equation}
V(t) = \frac {2}{5} \frac {R(t)}{t} 
\label{velocity}
\quad ,
\end{equation}
which  expressed in astrophysical units 
is 
\begin{equation}
V (t) \approx  
7.746 \,{\frac {\sqrt [5]{{\it E_{42}}}}{\sqrt [5]{{\it n_0}}{{\it 
t_4}}^{3/5}}}~  \frac{km}{s} 
\quad . 
\label{velocityastro}
\end {equation}
By inserting  
$M_{\sun}$=0.605 and  $R_{pc}$=1
in  formula~(\ref{concagb})
 we obtain
$n \approx 6.28 \frac{particles}{cm^3} $ .
This value is higher than the value of number density
of the  ISM at the plane of the galaxy, 
$n \approx1  \frac{particles}{cm^3}$ .
Equations~(\ref{sedovastro})  and (\ref{velocityastro})  
represent a system of two equations in two 
unknowns : $t_4$ and  $E_{42}$ . By
inserting for example  $R=0.78~pc$ 
in equation~(\ref{sedovastro})
we find 
\begin{equation}
t_4= 77.15\,{\frac {1}{\sqrt {{\it E_{42}}}}}
\label{t4equation}
\quad  ,
\end{equation}
and    inserting   
$V=35~km\;s^{-1}$
in equation~(\ref{velocityastro})
we obtain  
\begin{equation}
0.3954\,\sqrt {{\it E_{42}}}=35
\quad .
\end{equation}
The previous equation is solved for   
$E_{42} = 7833.4$ 
that according to equation~(\ref{t4equation}) 
means $t_4$=.87173.
These two parameters allows a rough evaluation
of the mechanical luminosity $L=\frac{E}{t}$ 
that turns out to be $L\approx 2.847 \;10^{34} ergs\;s^{-1} $.
This value should be  bigger than  the observed luminosities 
in the various bands.
As an example the X-ray  luminosity of PNs , $L_X$, in the wavelength band 
5-28~\AA~  
has a range $[10^{30.9} \leftrightarrow  10^{31.2} ergs\;s^{-1}  ]$  
, see Table 3 in \cite{Steffen2008}. 

Due to the fact  that is difficult to compute the 
volume  in an asymmetric expansion the Sedov solution
is adopted only in this paragraph.


\subsection{Spherical Symmetry - Momentum Conservation}

The thin layer approximation assumes that all the swept-up 
gas accumulates infinitely in a thin shell just after
the shock front.
The conservation of the radial momentum requires that 
\begin{equation}
\frac{4}{3} \pi R^3 \rho \dot {R} = M_0
\quad ,
\end{equation}
where $R$ and $\dot{R}$   are  the radius and the velocity
of the advancing shock ,
$\rho$ the density of the ambient medium ,
$M_0$ the momentum evaluated at $t=t_0$ ,
$R_0$ the initial radius  
and 
$\dot {R_0}$  the  initial velocity ,
see \cite{Dyson1997,Padmanabhan_II_2001}.
The law of motion is 
\begin{equation}
R = R_0 \left  ( 1 +4 \frac{\dot {R_0}} {R_0}(t-t_0) \right )^{\frac{1}{4}}  
\label{radiusm}
\quad .
\end{equation}  
and the velocity 
\begin{equation}
\dot {R} = \dot {R_0} \left ( 1 +4 \frac{\dot {R_0}} {R_0}(t-t_0)\right )^{-\frac{3}{4}}  
\label{velocitym} 
\quad . 
\end{equation}   
From equation (\ref{radiusm}) we can extract $\dot {R_0}$ 
and insert it in equation (\ref{velocitym})
\begin{equation}
\dot {R} =\frac{1}{4(t-t_0)}  \frac{R^4-R_0^4}{R_0^3}  
\left  ( 1+\frac{R^4-R_0^4}{R_0^4} \right )^{-\frac{3}{4}} 
\label{velocitym2} 
\quad .
\end{equation}
The astrophysical  units are:  $t_4$  and 
$t_{0,4}$ 
which  are $t$ and  $t_0$ 
expressed  in $10^4$ \mbox{yr} units, 
$R_{pc}$ and $R_{0,pc}$ which are  
$R$ and  $R_0$  expressed in  $pc$,
$\dot {R}_{kms}$ 
and
$\dot {R}_{0,kms}$ 
which are 
 $\dot{R} $ and  $\dot{R}_0$   expressed 
in $\frac{km}{s}$.
Therefore the previous formula becomes 
\begin{equation}
\dot {R}_{kms} =24.49 \frac{1}{(t_4-t_{0,4} )}  
\frac{R_{pc}^4-R_{0,pc}^4}
{R_{0,pc} ^3}  
\left  ( 1+\frac{R_{pc}^4-R_{0,pc}^4}{R_{0,pc}^4} \right )^{-\frac{3}{4}} 
\label{velocitym2astro} 
\quad .
\end{equation}
On introducing   $R_{0,pc}=0.1$ , 
$R_{pc}=0.78$ ,                
$\dot {R}_{kms}  =  34.5  \frac{km}{s}$ ,
the approximated age of A39  is  found 
to be $t_4-t_{0,4}=50 $ and $\dot {R}_{0,kms}  =  181.2 $.  

\subsection{Asymmetry - Momentum Conservation}

Given the Cartesian   coordinate system
$(x,y,z)$ ,
the plane $z=0$ will be called equatorial plane 
and in  polar coordinates $z= R \sin ( \theta) $,
 where
$\theta$ is the polar angle and $R$ the distance from 
the origin .
The presence of a non homogeneous medium in which the expansion 
takes place can be modeled assuming an exponential 
behavior for the number of particles of the type 
\begin{equation}
n (z) = n_0 \exp {- \frac {z}{h} } 
\quad  
= n_0 \exp {- \frac {R\times \sin (\theta) }{h} } 
\quad  , 
\end{equation}
where  $R$ is the radius of the shell , $n_0$ is the number
of particles at $R=R_0$ and $h$ the scale.
The 3D expansion will be characterized by the following 
properties 
\begin {itemize}
\item Dependence of the momentary radius of the shell 
      on  the polar angle $\theta$ that has a range 
      $[-90 ^{\circ}  \leftrightarrow  +90 ^{\circ} ]$.

\item Independence of the momentary radius of the shell 
      from  $\phi$ , the azimuthal  angle  in the x-y  plane,
      that has a range  
      $[0 ^{\circ}  \leftrightarrow  360 ^{\circ} ]$.
\end {itemize}
The mass swept, $M$,  along the solid angle
$ \Delta\;\Omega $,  between 0 and $R$ is
\begin{equation}
M(R)= 
\frac { \Delta\;\Omega } {3}  m_H n_0 I_m(R) 
+ \frac{4}{3} \pi R_0^3 n_0 m_H
\quad  , 
\end {equation}
where 
\begin{equation}
I_m(R)  = \int_{R_0} ^R r^2 \exp { - \frac {r \sin (\theta) }{ h}  } dr
\quad ,
\end{equation}
where $R_0$ is the initial radius  
and $m_H$ the mass of the hydrogen  .
The integral is   
\begin{eqnarray}
I_m(R)  =
\frac
{
h \left( 2\,{h}^{2}+2\,R_0h\sin \left( \theta \right) +{R_0}^{2} \left( 
\sin \left( \theta \right)  \right) ^{2} \right) {{\rm e}^{-{\frac {R_0
\sin \left( \theta \right) }{h}}}}
}
{
\left( \sin \left( \theta \right)  \right) ^{3}
}
           \nonumber\\
- \frac
{
h \left( 2\,{h}^{2}+2\,Rh\sin \left( \theta \right) +{R}^{2} \left( 
\sin \left( \theta \right)  \right) ^{2} \right) {{\rm e}^{-{\frac {R
\sin \left( \theta \right) }{h}}}}
}
{
\left( \sin \left( \theta \right)  \right) ^{3}
}
\quad .
\end{eqnarray}
The conservation of the momentum gives 
\begin{equation}
M(R)   \dot {R}= 
M(R_0) \dot {R_0}
\quad  ,
\end{equation}
where $\dot {R}$  is the  velocity 
at $R$ and 
$\dot {R_0}$  the  initial velocity at $R=R_0$.

In this differential equation of the first order in $R$ the 
variable can be separated and the integration 
term by term gives 
\begin{equation}
\int_{R_0}^{R}  M(r)  dr = 
M(R_0) \dot {R_0} \times ( t-t_0) 
\quad  ,
\end{equation}
where  $t$ is the time and $t_0$ the time at $R_0$.
The resulting non linear equation ${\mathcal{F}}_{NL}$ 
expressed in astrophysical units 
is 
\begin{eqnarray}
{\mathcal{F}}_{NL} = 
- 6\,{{\rm e}^{- \,{\frac {{\it R_{0,pc}}\,\sin \left( \theta
 \right) }{{\it h_{pc}}}}}}{{\it h_{pc}}}^{4}
- \,{\it h_{pc}}\,{{\rm e}^{-
 \,{\frac {{\it R_{0,pc}}\,\sin \left( \theta \right) }{{\it h_{pc}}}}}}
 \left( \sin \left( \theta \right)  \right) ^{3}{{\it R_{0,pc}}}^{3}
\nonumber \\
- 6
\,{{\it h_{pc}}}^{3}{{\rm e}^{- \,{\frac {{\it R_{0,pc}}\,\sin \left( 
\theta \right) }{{\it h_{pc}}}}}}\sin \left( \theta \right) {\it R_{0,pc}}-
 3\,{{\it h_{pc}}}^{2}{{\rm e}^{- \,{\frac {{\it R_{0,pc}}\,\sin \left( 
\theta \right) }{{\it h_{pc}}}}}} \left( \sin \left( \theta \right) 
 \right) ^{2}{{\it R_{0,pc}}}^{2}
\nonumber \\
- \,{{\it R_{0,pc}}}^{4} \left( \sin
 \left( \theta \right)  \right) ^{4}+ 6\,{{\rm e}^{- \,{\frac {{
\it R_{pc}}\,\sin \left( \theta \right) }{{\it h_{pc}}}}}}{{\it h_{pc}}}^{4}+
 4\,{{\rm e}^{- \,{\frac {{\it R_{pc}}\,\sin \left( \theta \right) }
{{\it h_{pc}}}}}}{{\it h_{pc}}}^{3}{\it R_{pc}}\,\sin \left( \theta \right) 
\nonumber  \\
+{
{\rm e}^{- \,{\frac {{\it R_{pc}}\,\sin \left( \theta \right) }{{\it 
h_{pc}}}}}}{{\it h_{pc}}}^{2}{{\it R_{pc}}}^{2} \left( \sin \left( \theta
 \right)  \right) ^{2}
\nonumber  \\
+ 2\,{{\rm e}^{- \,{\frac {{\it R_{0,pc}}\,\sin
 \left( \theta \right) }{{\it h_{pc}}}}}}{{\it h_{pc}}}^{3}{\it R_{pc}}\,\sin
 \left( \theta \right) + 2\,{{\rm e}^{- \,{\frac {{\it R_{0,pc}}\,
\sin \left( \theta \right) }{{\it h_{pc}}}}}}{{\it h_{pc}}}^{2}{\it R_{pc}}\,
 \left( \sin \left( \theta \right)  \right) ^{2}{\it R_{0,pc}}
\nonumber \\
+{{\rm e}^{-
 \,{\frac {{\it R_{0,pc}}\,\sin \left( \theta \right) }{{\it h_{pc}}}}}}{
\it h_{pc}}\,{\it R_{pc}}\, \left( \sin \left( \theta \right)  \right) ^{3}{
{\it R_{0,pc}}}^{2}
\nonumber \\
+ \left( \sin \left( \theta \right)  \right) ^{4}{{\it 
R_{0,pc}}}^{3}{\it R_{pc}}- 0.01 \left( \sin \left( \theta
 \right)  \right) ^{4}{{\it R_{0,pc}}}^{3}{ \dot{R}_{0,kms}}\, 
\left( t_4-t_{0,4}
 \right) 
=0  
\quad  ,
\end{eqnarray}
where   $t_4$  and 
$t_{0,4}$ 
 are $t$ and  $t_0$ 
expressed  in $10^4$ \mbox{yr} units, 
$R_{pc}$ and $R_{0,pc}$  are  
$R$ and  $R_0$  expressed in  $pc$,
$\dot {R}_{kms}$ 
and
$\dot {R}_{0,kms}$ 
are 
$\dot{R} $ and  $\dot{R}_0$   expressed 
in $\frac{km}{s}$,
$\theta$ is expressed in radians     
and  $h_{pc}$ is  the  the scale , $h$  ,
expressed in $pc$.
It is not possible to find  $R_{pc}$   analytically  and
a numerical method   should be implemented.
In  our case in order  
to find  the root of  ${\mathcal{F}}_{NL}$,
the FORTRAN SUBROUTINE  ZRIDDR from \cite{press} has been used.

The unknown parameter $t_4-t_{0,4}$  can be found 
from different runs  of the code once
 $R_{0,pc}$ is fixed as $\approx$ 1/10 of the observed equatorial
radius , $\dot {R}_{0,kms}$ is  200  or less
and   
$h_{pc} \approx 2\times R_{0,pc}$.

From a practical point  of view,  
$\epsilon$ ,
the percentage  of
reliability  of our code can also be  introduced,
\begin{equation}
\epsilon  =(1- \frac{\vert( R_{\mathrm {pc,obs}}- R_{pc,\mathrm {num}}) \vert}
{R_{pc,\mathrm {obs}}}) \cdot 100
\,,
\label{efficiency}
\end{equation}
where $R_{pc,\mathrm {obs}}$ is the   radius as given 
by the astronomical observations in parsec ,
and  $R_{pc,\mathrm {num}}$ the radius  obtained from our  simulation
in parsec.

In order to  test the  simulation over different angles,  an
observational percentage  of 
reliability 
,$\epsilon_{\mathrm {obs}}$,
is  introduced which  uses 
both the size and the shape, 
\begin{equation}
\epsilon_{\mathrm {obs}}  =100(1-\frac{\sum_j |R_{pc,\mathrm {obs}}-R_{pc,\mathrm {num}}|_j}{\sum_j
{R_{pc,\mathrm {obs}}}_{,j}})
, 
\label{efficiencymany}  
\end{equation}
where 
the  index $j$  varies  from 1 to the number of
available observations.

\subsubsection{Simulation of the Ring nebula}

A typical set of parameters that allows
us  to simulate
the Ring nebula is reported in 
Table~\ref{parameters}.

\begin{table}
      \caption{Data of the simulation of the Ring nebula }
         \label{parameters}
      \[
         \begin{array}{cc}
            \hline
            \noalign{\smallskip}
\mbox {Initial ~expansion~velocity~,${\dot R}_{{0,kms}}$  } & 200             \\
\mbox {Age~($t_4-t_{0,4}$)  }                     & 0.12 \\
\mbox {Initial~radius~ $R_{0,pc}$  }                        & 0.035 \\
\mbox {scaling~ $h_{pc}$    }                                & \mbox {$2\times R_{0,pc}$}\\
            \noalign{\smallskip}
            \hline
         \end{array}
      \]
   \end{table}

\begin{figure}
  \begin{center}
\includegraphics[width=8cm]{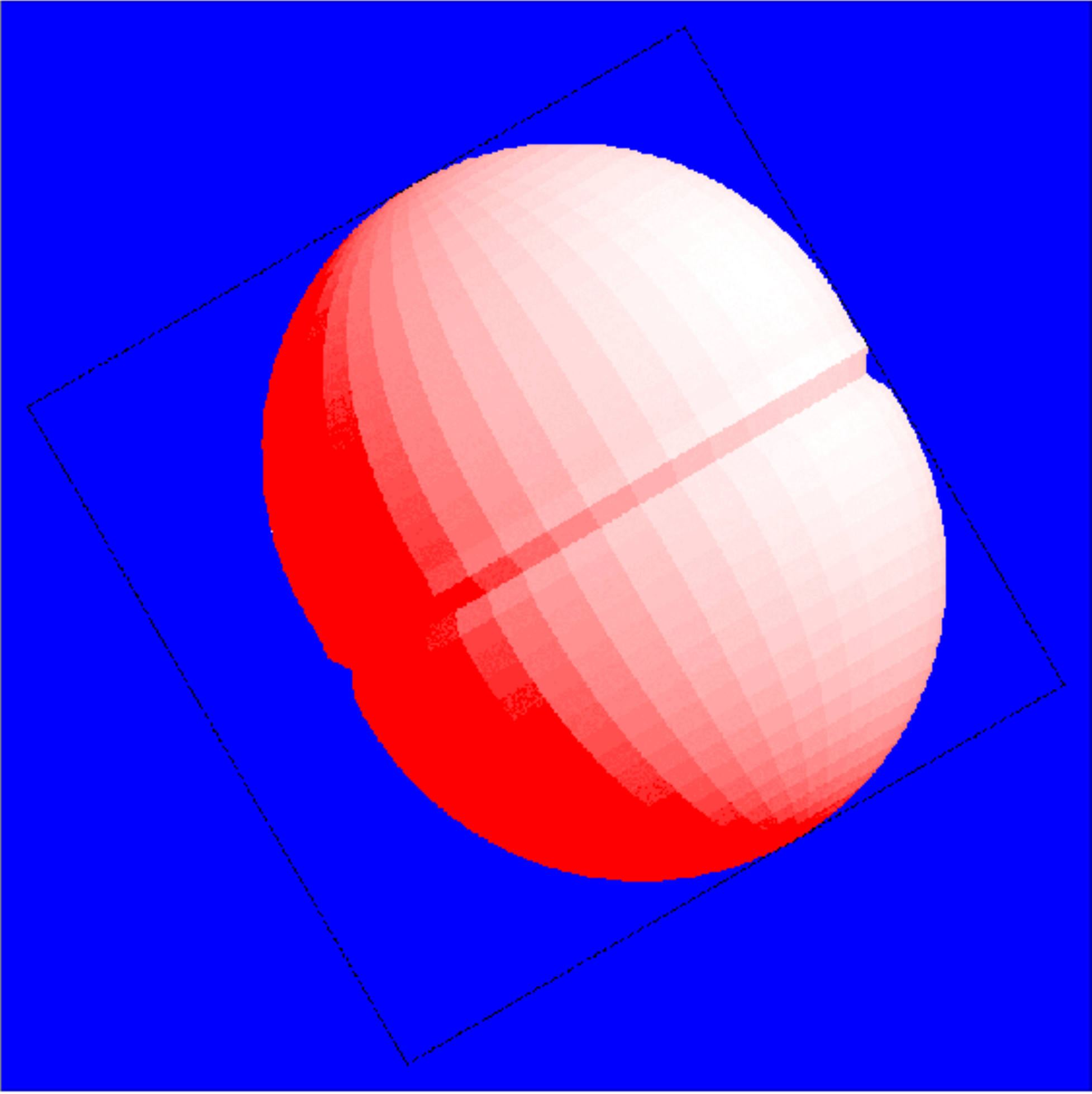}
  \end {center}
\caption {
Continuous  three-dimensional surface of the Ring nebula :
the three Eulerian angles 
characterizing the point of view are 
     $ \Phi $=180    $^{\circ }  $, 
     $ \Theta $=90   $^{\circ }$
and  $ \Psi $=-30    $^{\circ }   $.
Physical parameters as in Table~\ref{parameters}.
          }%
    \label{ring_faces}
    \end{figure}

\begin{figure}
  \begin{center}
\includegraphics[width=8cm]{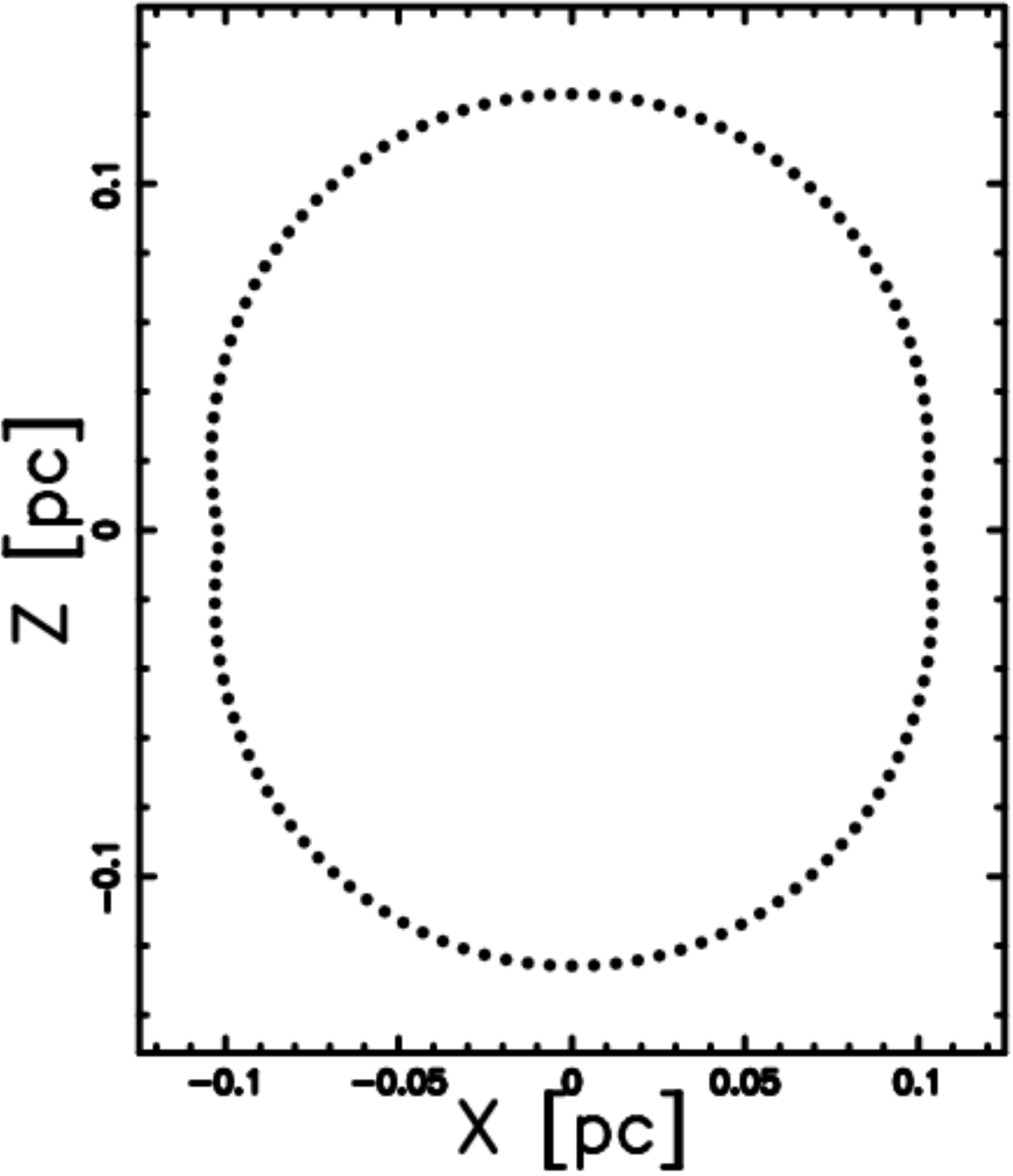}
  \end {center}
\caption {
Section of the Ring nebula  on the {\it x-z}  plane.
The horizontal and vertical axis are in $pc$.
Physical parameters as in Table~\ref{parameters}.
          }%
    \label{ring_cut}
    \end{figure}

\begin{figure}
  \begin{center}
\includegraphics[width=8cm]{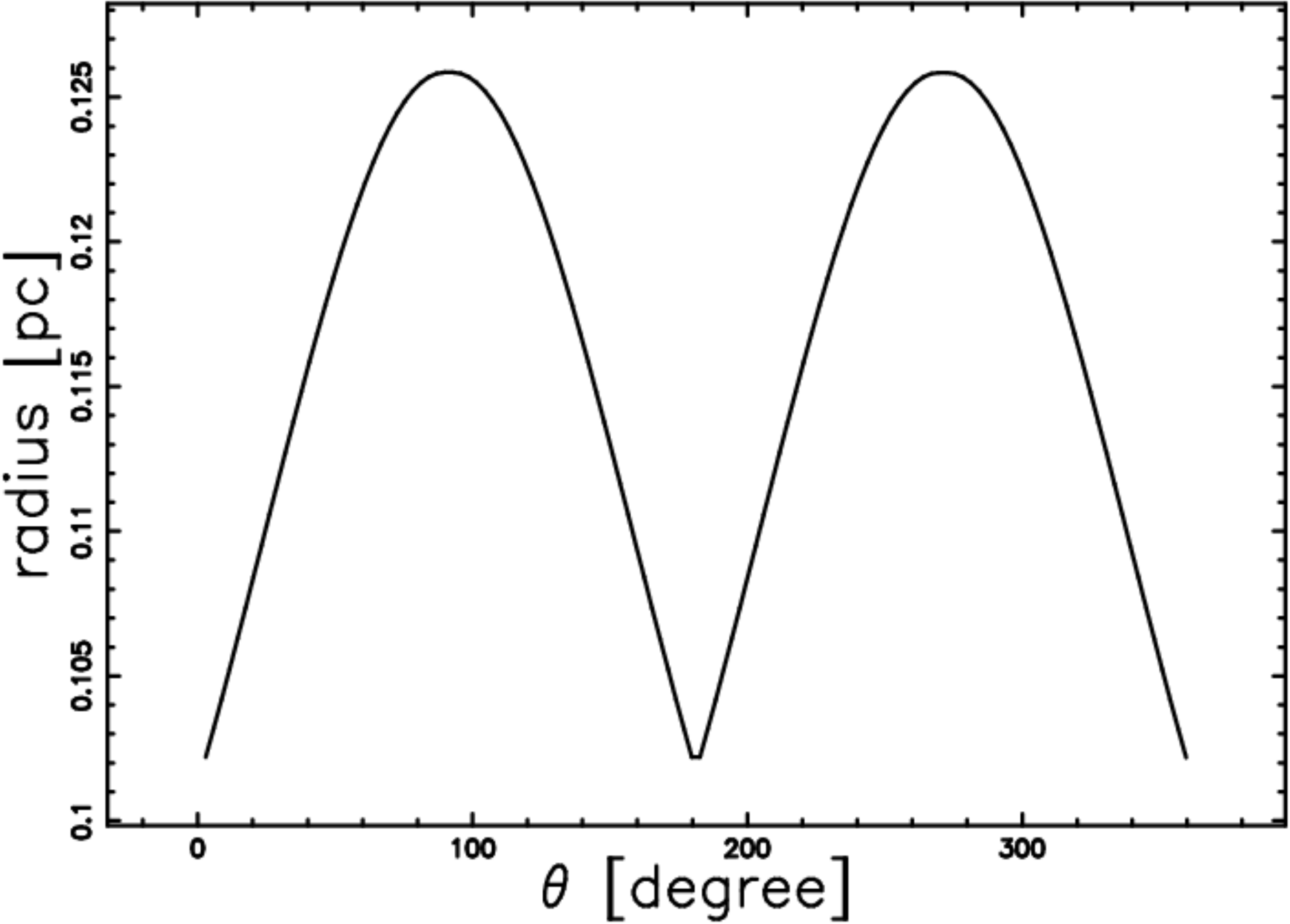}
  \end {center}
\caption {
Radius in $pc$  of the Ring nebula  
as a function of the position angle
in degrees.
Physical parameters as in Table~\ref{parameters}.
          }%
    \label{ring_radius}
    \end{figure}

\begin{figure}
  \begin{center}
\includegraphics[width=8cm]{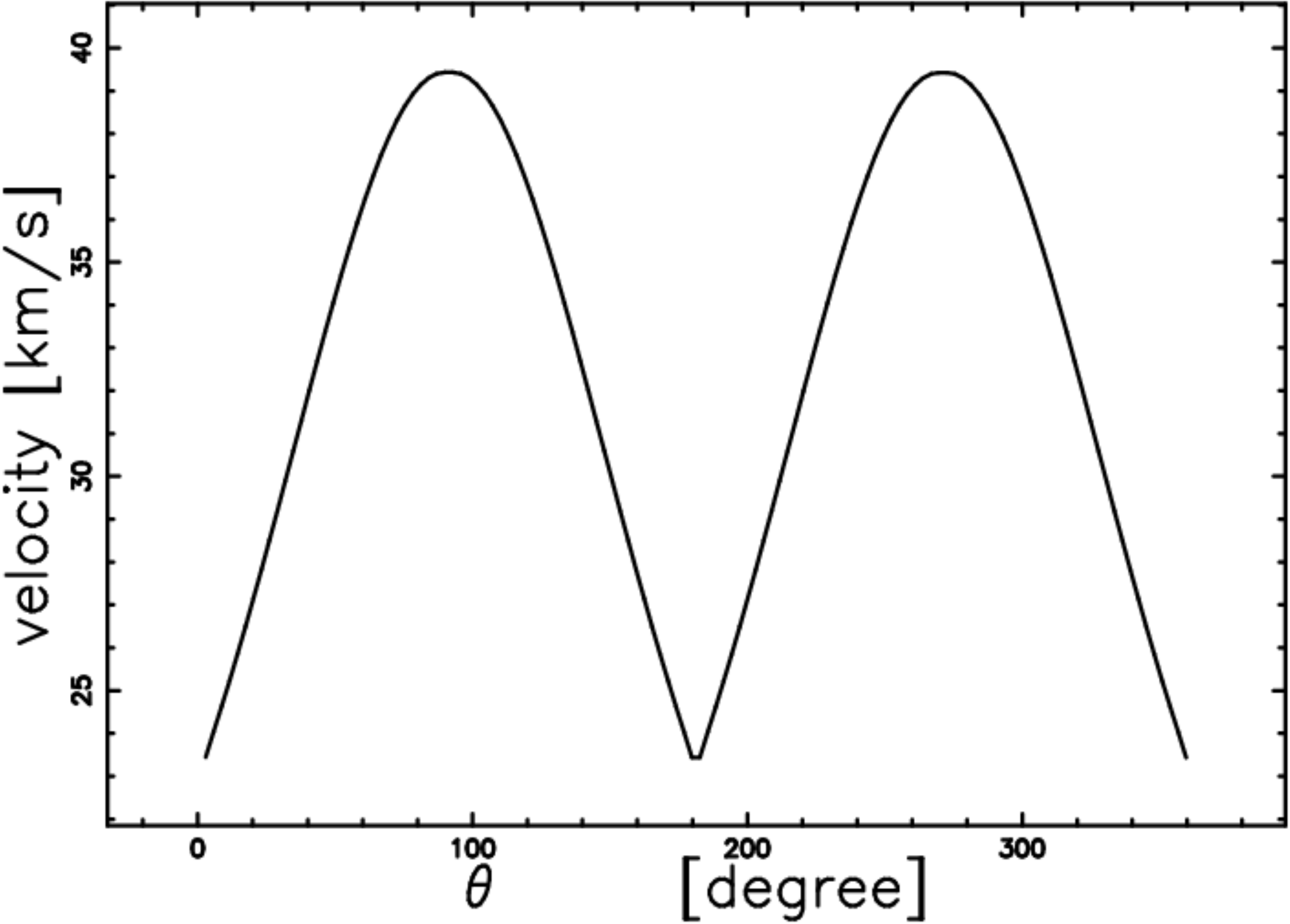}
  \end {center}
\caption {
Velocity  in $\frac{km}{s}$  of the Ring nebula  
as a  function of the position angle
in degrees.
Physical parameters as in Table~\ref{parameters}.
          }%
    \label{ring_velocity}
    \end{figure}

\begin{figure}
  \begin{center}
\includegraphics[width=8cm]{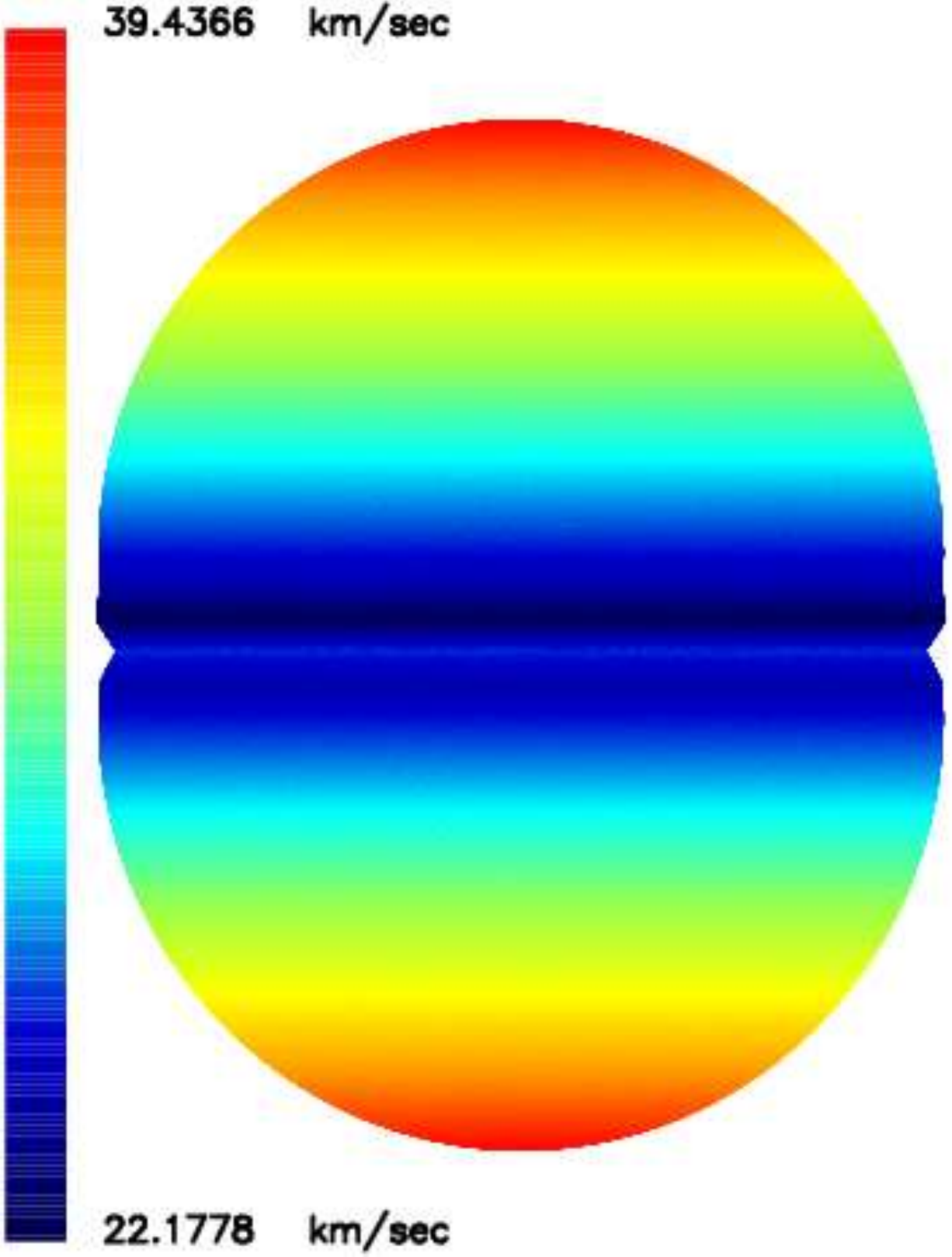}
  \end {center}
\caption {
  Map of the expansion velocity   in $\frac{km}{s}$
  relative to the simulation of the Ring nebula  
  when 300000 random points are selected on the surface.
  Physical parameters as in Table~\ref{parameters}.
          }%
    \label{ring_velocity_field}
    \end{figure}

The complex 3D behavior of the advancing Ring nebula  is reported 
in Figure~\ref{ring_faces} and Figure~\ref{ring_cut}
reports the asymmetric expansion in  a  section crossing the center.
In order to better visualize the asymmetries 
Figure~\ref{ring_radius} and  Figure~\ref{ring_velocity}
report the radius  and the velocity
as a function of the position angle $\theta$.
The combined effect of spatial asymmetry and field of velocity 
are reported in Figure \ref{ring_velocity_field}.

The efficiency  of our code in  reproducing
  the observed radii 
as given by formula~(\ref{efficiency} )
and  the efficiency when the age is five time greater 
are  reported in 
Table~\ref{tab:rad}.

An analogous formula  allows us to compute the efficiency
in the computation of the maximum velocity ,  
see Table~\ref{tab:vel}. 

\begin{table}
      \caption{Reliability of the radii of the Ring nebula.}
         \label{tab:rad}
      \[
         \begin{array}{lccc}
            \hline
            \noalign{\smallskip}
~~~~     & R_{\mathrm{up}}(\mathrm{pc})  ~polar~direction  &
           R_{\mathrm{eq}}  (\mathrm{pc})~equatorial~plane  \\
            \noalign{\smallskip}
            \hline
            \noalign{\smallskip}
R_{\mathrm {obs}}                       &  0.14    & 0.1        \\
R_{\mathrm {num}}   (\mbox{our~code})   &  0.125   & 0.102
        \\
\mbox {$\epsilon$} (\%)      &  89   &   97       \\
\mbox {$\epsilon$}~(\%)~for~a~time~5~times~greater       &  27   &   41       \\
            \noalign{\smallskip}
            \hline
         \end{array}
      \]
   \end{table}

\begin{table}
      \caption{Reliability of the velocity
          of the Ring nebula
        }
         \label{tab:vel}
      \[
         \begin{array}{lc}
            \hline
            \noalign{\smallskip}
~~~~     & V  ({\frac {km}{s}  })  ~maximum~velocity\\
            \noalign{\smallskip}
            \hline
            \noalign{\smallskip}
V_{\mathrm {obs}}          &  48.79       \\
V_{\mathrm {num}}          &  39.43       \\
\mbox {$\epsilon$} (\%)    &  80.81       \\
\mbox {$\epsilon$} (\%) ~(\%)~for~a~time~5~times~greater   &  35.67      \\
            \noalign{\smallskip}
            \hline
         \end{array}
      \]
   \end{table}

\subsubsection{Simulation of MyCn 18  }

A typical set of parameters that allows us  to simulate
  MyCn 18    is reported in 
Table~\ref{parametershom}.

\begin{table}
      \caption{Data of the simulation of MyCn 18 }
         \label{parametershom}
      \[
         \begin{array}{cc}
            \hline
            \noalign{\smallskip}
\mbox {Initial ~expansion~velocity~,${\dot R}_{{0,kms}}$ 
 [km~s$^{-1}$}] &  200             \\
\mbox {Age~($t_4-t_{0,4}$) [10$^4$~yr]}                    & 0.2   \\
\mbox {Initial~radius~ $R_{0,pc}$ ~[pc] }                       & 0.001 \\
\mbox {scaling~ h     [pc] }                               & \mbox{$1.0\times R_0$} \\
            \noalign{\smallskip}
            \hline
         \end{array}
      \]
   \end{table}

\begin{figure}
  \begin{center}
\includegraphics[width=8cm]{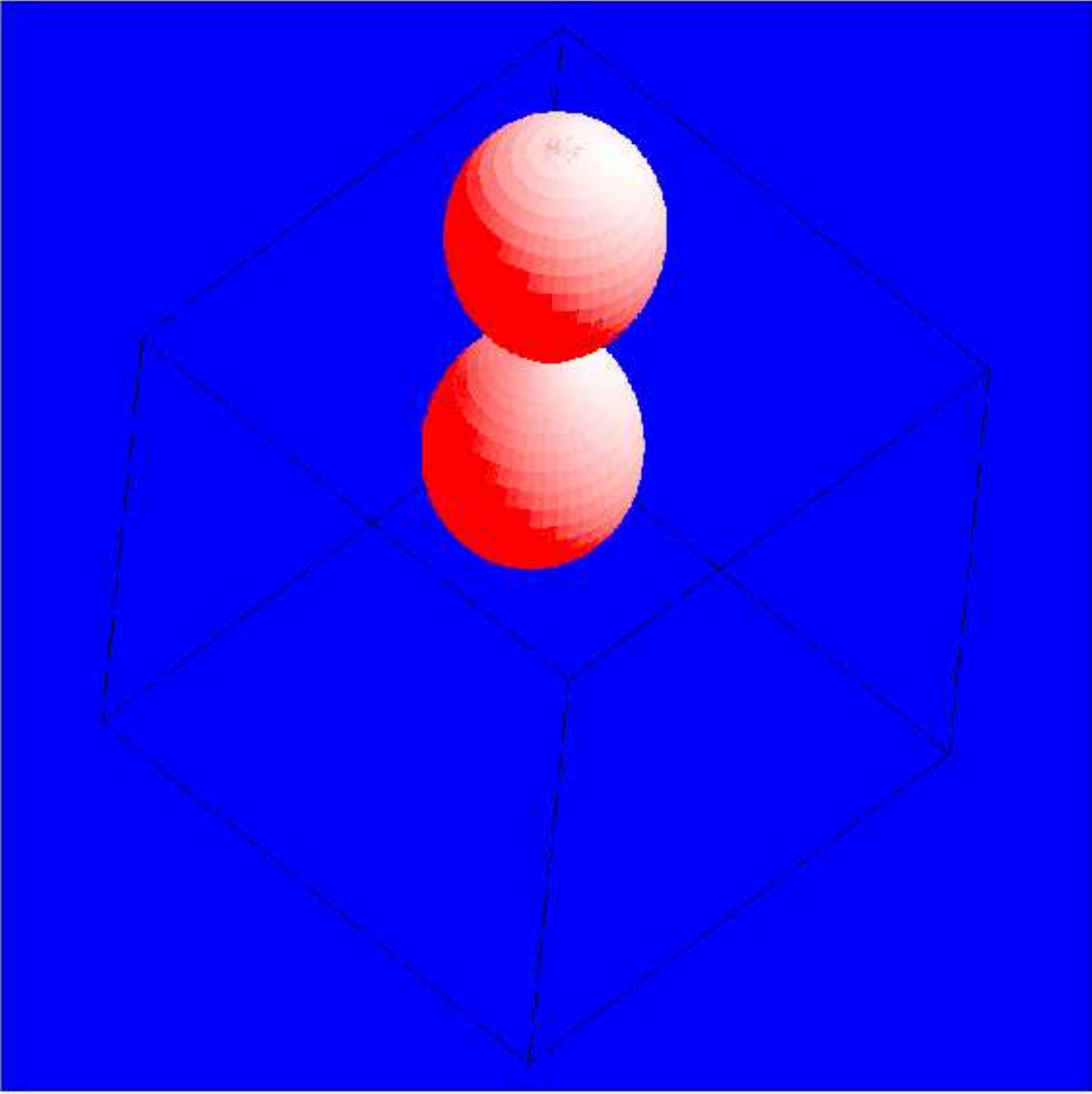}
  \end {center}
\caption {
Continuous  three-dimensional surface of MyCn 18 :
the three Eulerian angles 
characterizing the point of view are 
     $ \Phi   $=130     $^{\circ }  $, 
     $ \Theta $=40   $^{\circ }$
and  $ \Psi   $=5     $^{\circ }   $.
Physical parameters as in Table~\ref{parametershom}.
          }%
    \label{mycn18_faces}
    \end{figure}

\begin{figure}
  \begin{center}
\includegraphics[width=8cm]{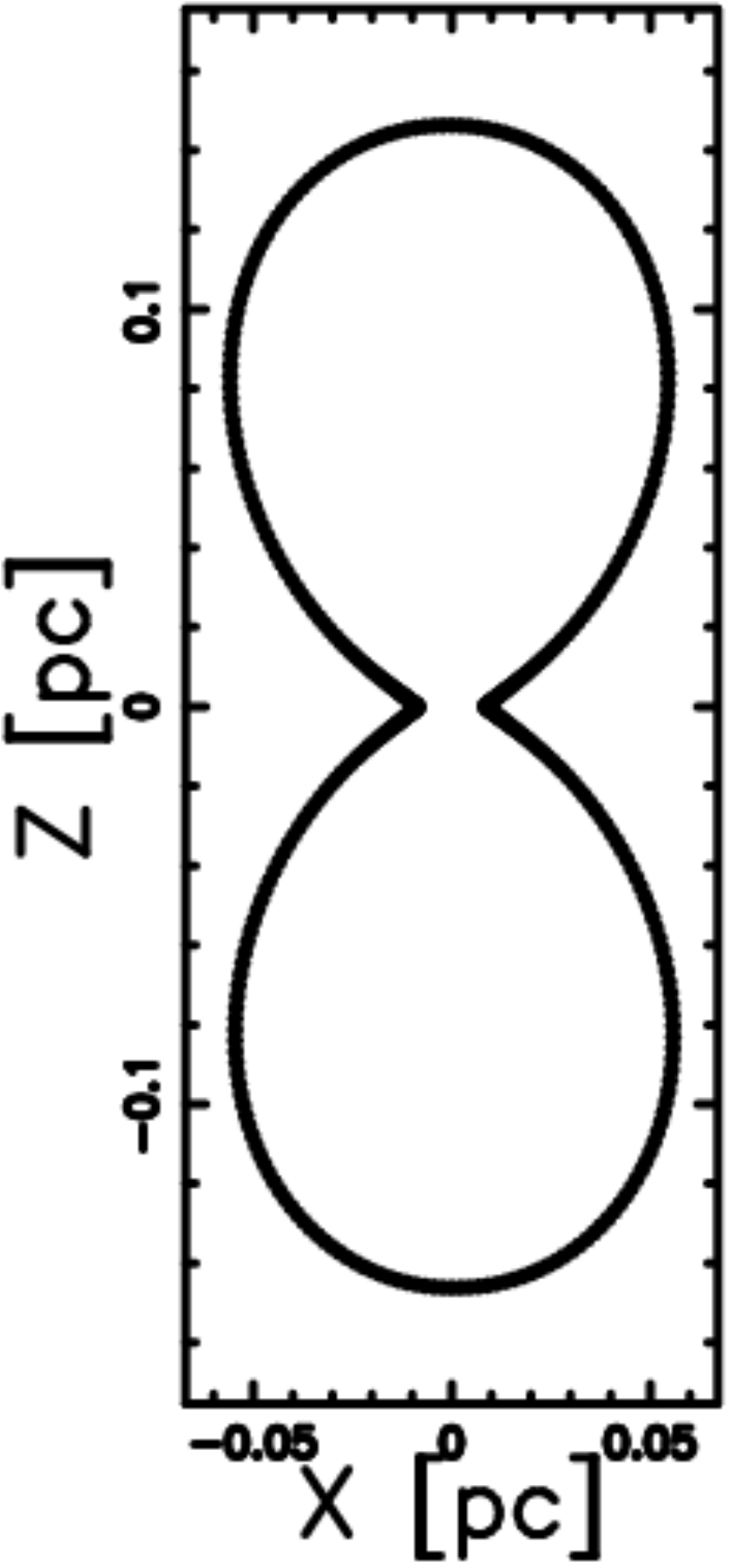}
  \end {center}
\caption {
Section of MyCn 18  on the {\it x-z}  plane.
Physical parameters as in Table~\ref{parametershom}.
          }%
    \label{mycn18_cut}
    \end{figure}

\begin{figure}
  \begin{center}
\includegraphics[width=8cm]{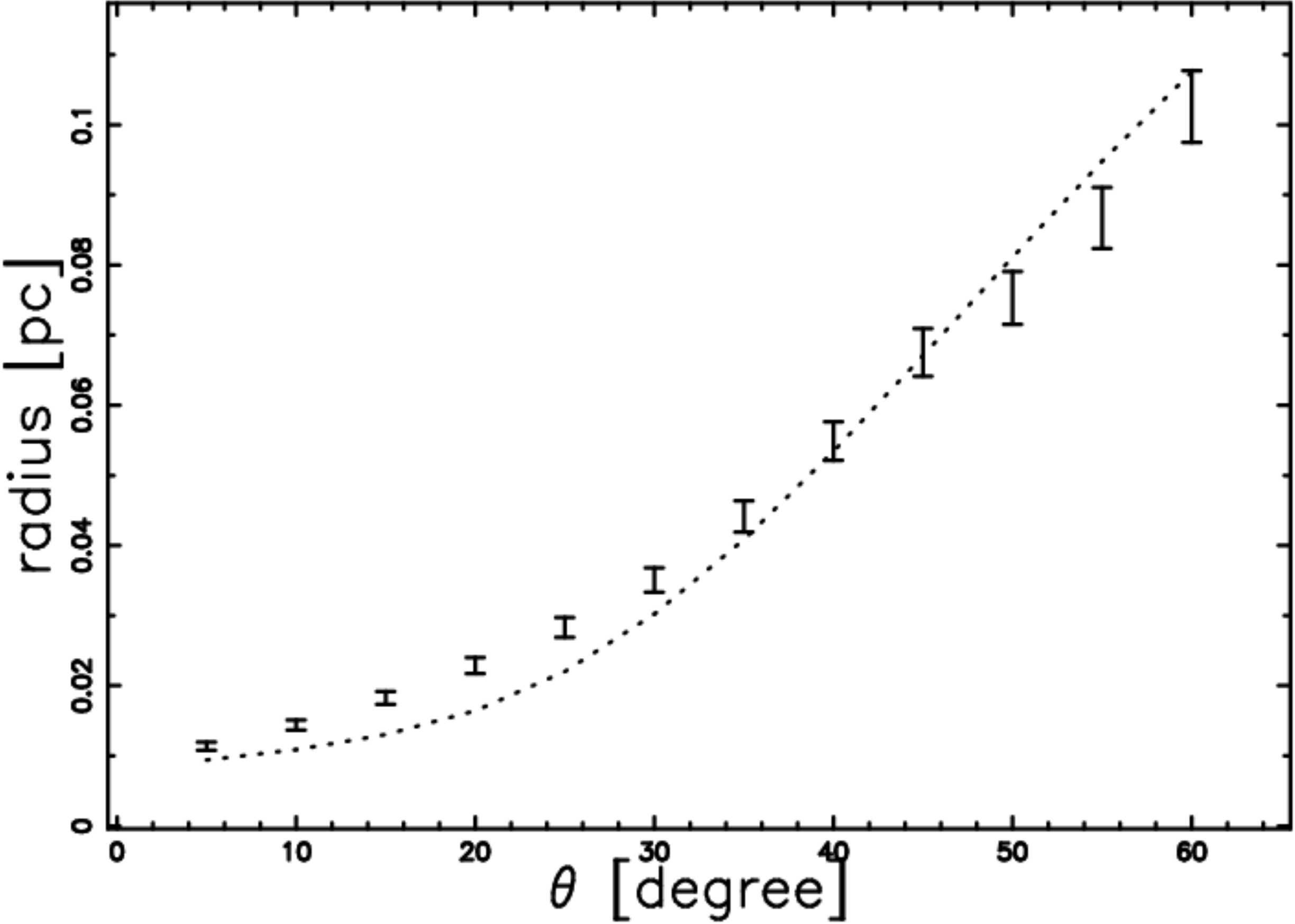}
  \end {center}
\caption {
Radius in pc  of  MyCn 18  
as a function of latitude 
from  $0^{\circ}$  to $60^{\circ}$ ( dotted line) 
when the physical parameters are those of   Table~\ref{parametershom}.
The points with error bar (1/10 of the value) 
represent the data of 
Table 1 in  Dayal et al. 2000. 
          }%
    \label{mycn18_radius}
    \end{figure}

The bipolar  behavior of the advancing 
MyCn 18 
 is reported 
in Figure~\ref{mycn18_faces} and Figure~\ref{mycn18_cut}
reports the expansion in  a section crossing the center.
It is   interesting to point out  the  similarities 
between our Figure~\ref{mycn18_cut} of  
MyCn 18 
and Figure~1 in  \cite{Morisset2008} 
which define the parameters $a$ and $h$ 
of  the Atlas of synthetic line profiles.
In order to better visualize the two lobes  
Figure~\ref{mycn18_radius} 
reports  the radius 
as a function of the position angle $\theta$.

\begin{figure}
  \begin{center}
\includegraphics[width=5cm]{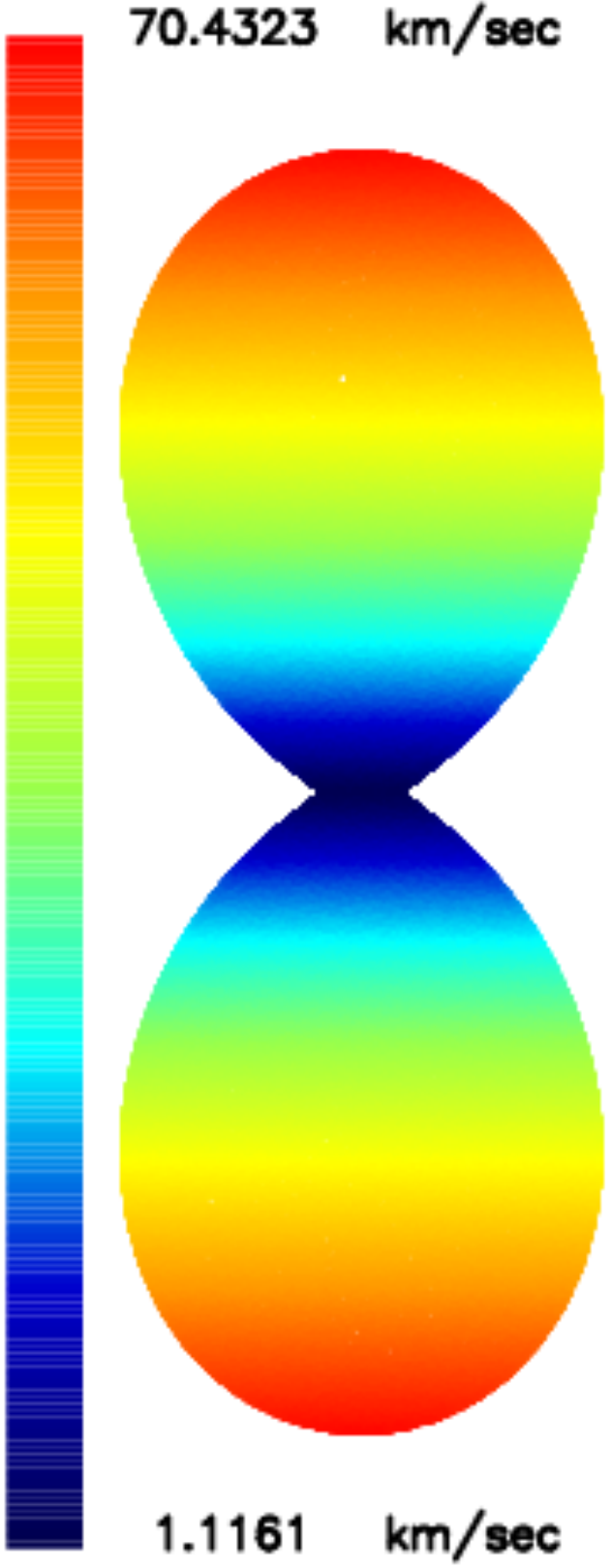}
  \end {center}
\caption {
  Map of the expansion velocity   in $\frac{km}{s}$
  relative to the simulation of MyCn 18  
  when 300000 random points are selected on the surface.
  Physical parameters as in Table~\ref{parametershom}.
          }%
    \label{mycn18_velocity_field}
    \end{figure}

The combined effect of spatial asymmetry and field of velocity 
are reported in Figure \ref{mycn18_velocity_field}.

The efficiency  of our code in  reproducing   the spatial 
shape  over 12 directions of  MyCn 18
as given by formula~(\ref{efficiencymany} ) is reported in 
Table~\ref{tab:radmycn18}. 
This Table also reports the efficiency in simulating the
shape of the velocity.

\begin{figure}
  \begin{center}
\includegraphics[width=8cm]{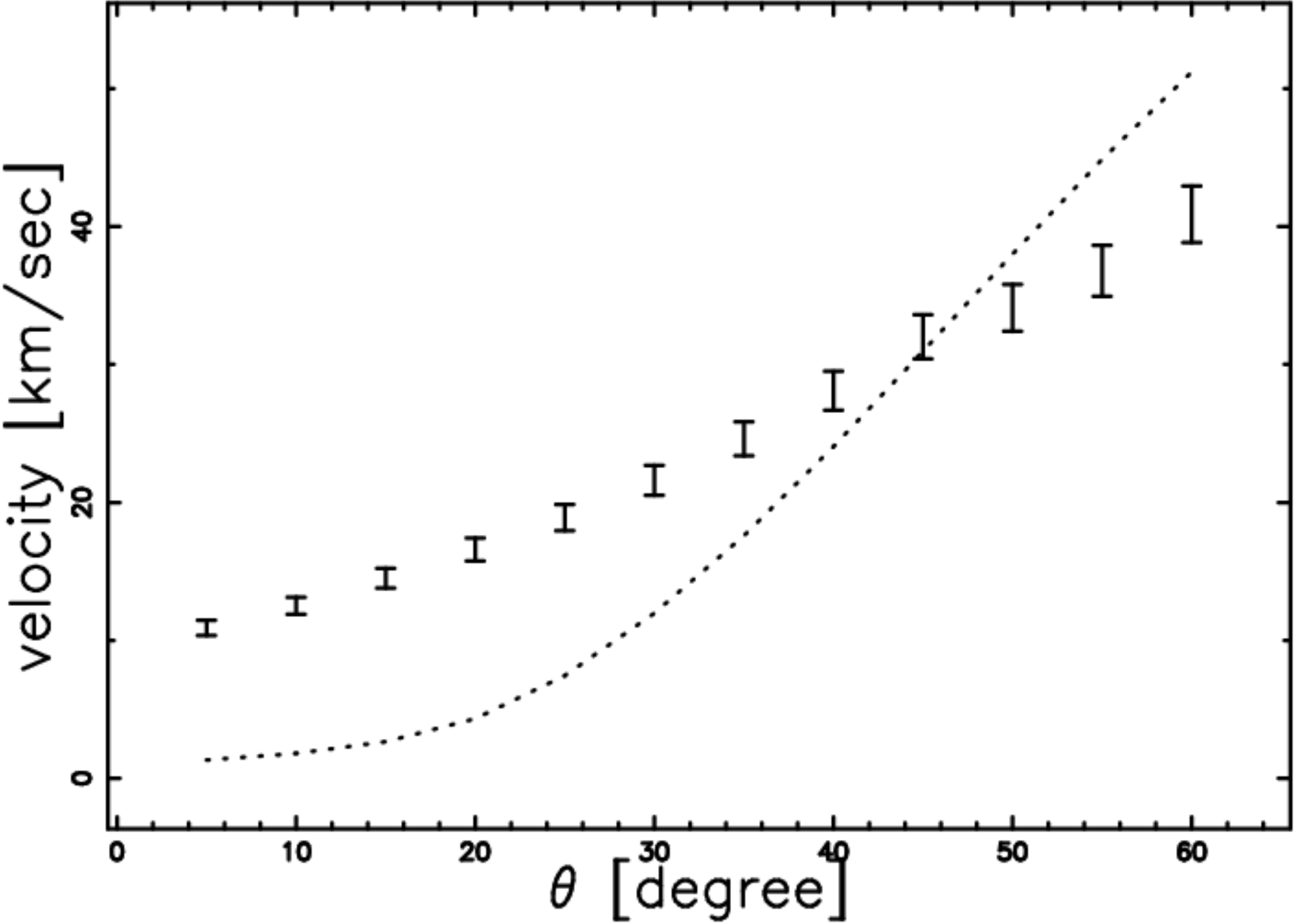}
  \end {center}
\caption {
Velocity  in $\frac{km}{s}$  of MyCn 18    
as a  function of the latitude in degrees  
when  the physical parameters  are those of 
 Table~\ref{parametershom}, dotted  line.
The points with error bar (1/10 of the value) 
represent the data of 
Table 1 in  Dayal et al. 2000. 
          }%
    \label{mycn18_velocity}
    \end{figure}

\begin{table}
      \caption{Reliability of the spatial and velocity
              shape of MyCn 18.}
         \label{tab:radmycn18}
      \[
         \begin{array}{lcc}
            \hline
            \noalign{\smallskip}
~~~                                & radius   &  velocity    \\
\mbox {$\epsilon_{obs}$} (\%)      &  90.66   &  57.68        \\
            \noalign{\smallskip}
            \hline
         \end{array}
      \]
   \end{table}
Figure ~ \ref{mycn18_velocity} reports our results 
as well those of Table~1 in \cite{Dayal2000}.

\section{Diffusion}
\label{sec_diffusion}

The mathematical diffusion  allows us  to  follow the number density
of particles from high values
(injection) to low values (absorption).
We recall that the number density is expressed
in $\frac{particles}{unit~ volume}$ and  the symbol $C$ is 
used  in the mathematical diffusion and 
the symbol $n$ in an astrophysical context.
The density $\rho$ is obtained 
by 
multiplying $n$ 
by the mass of  hydrogen , $m_H$ ,
 and by a multiplicative factor , $f$,
which varies from 1.27 in   \cite{Kim2000} to 1.4 
in \cite{Dalgarno1987}
\begin{equation}
\rho  = f  m_H n 
\quad .
\end{equation}
The physical process that allows  the particles
to diffuse  is hidden in the mathematical diffusion.
In our case the physical process can be the random walk 
with a time step equal  to the Larmor gyroradius.
In the Monte Carlo diffusion 
the step-length  of the random walk
is generally taken as  a fraction of the side of the 
considered box.
Both mathematical diffusion 
and Monte Carlo diffusion use the concept of absorbing-boundary  which is  the spatial coordinate where
the diffusion path  terminates.

In the following,  3D mathematical diffusion from a
sphere and 1D  mathematical as well  Monte Carlo diffusion
in presence of drift are considered.

\subsection{3D diffusion from a spherical source}
\label{mathematical}
Once the number density , $C$, and the diffusion coefficient ,$D$,
are introduced ,  Fick'~s first equation
changes   expression on the basis  of the adopted
environment  ,  see for example equation~(2.5) in  \cite{berg}.
In three dimensions  it is 
\begin{equation}
\frac {\partial C }{\partial t} =
D \nabla^2 C
\quad,
\label{eqfick}
\end {equation}
where $t$ is the time  and $ \nabla^2$ is 
the Laplacian differential operator.

In presence of the  steady state condition:
\begin{equation}
D \nabla^2 C   = 0
\quad .
\label{eqfick_steady}
\end {equation}

The  number density rises from 0 at {\it r=a}  to a
maximum value $C_m$ at {\it r=b} and then  falls again
to 0 at {\it  r=c}~.
The  solution to  equation~(\ref{eqfick_steady})
 is
\begin{equation}
C(r) = A +\frac {B}{r}
\quad,
\label{solution}
\end {equation}
where $A$ and $B$  are determined by  the boundary conditions~,
\begin{equation}
C_{ab}(r) =
C_{{m}} \left( 1-{\frac {a}{r}} \right)  \left( 1-{\frac {a}{b}}
 \right) ^{-1}
\quad a \leq r \leq b
\quad,
\label{cab} 
\end{equation}
and
\begin{equation}
C_{bc}(r)=
C_{{m}} \left( {\frac {c}{r}}-1 \right)  \left( {\frac {c}{b}}-1
 \right) ^{-1}
\quad b \leq r \leq c
\quad.
\label{cbc}
\end{equation}
These solutions can be found in 
\cite{berg}
or in 
\cite{crank} .

\subsection{1D diffusion with drift, mathematical diffusion}

In one dimension  and in the presence of a drift velocity
,$u$,
along the  radial direction 
the diffusion is governed by  Fick's second equation ,
see equation~(4.5) in \cite{berg} ,
\begin{equation}
\frac {\partial C }{\partial t} =
D  \frac {\partial^2C}{\partial r^2} -  {\vec{u}} \frac {\partial C}{\partial r}
\quad ,
\label{eqfick_1_drift}
\end {equation}
where ${\vec{u}}$  can take  two directions.
The  number density rises from 0 at {\it r=a}  to a
maximum value $C_m$ at {\it r=b} and then  falls again
to 0 at {\it  r=c}~.
The general solution to  equation~(\ref{eqfick_1_drift})
in presence of a steady state is 
\begin{equation}
C(r) = A + B e^{{\frac{\vec {u}}{D}}r } 
\quad. 
\end{equation}
We now assume  that   {\it u}  and {\it r} 
do not  have  the same  direction  and therefore 
{\it u } is negative ;  the solution is
\begin{equation}
C(r) = A + B e^{-{\frac{u}{D}}r } 
\quad ,
\label{solution_1D_drift}
\end{equation}
and now the velocity $u$ is a scalar.

\begin{figure}
  \begin{center}
\includegraphics[width=8cm]{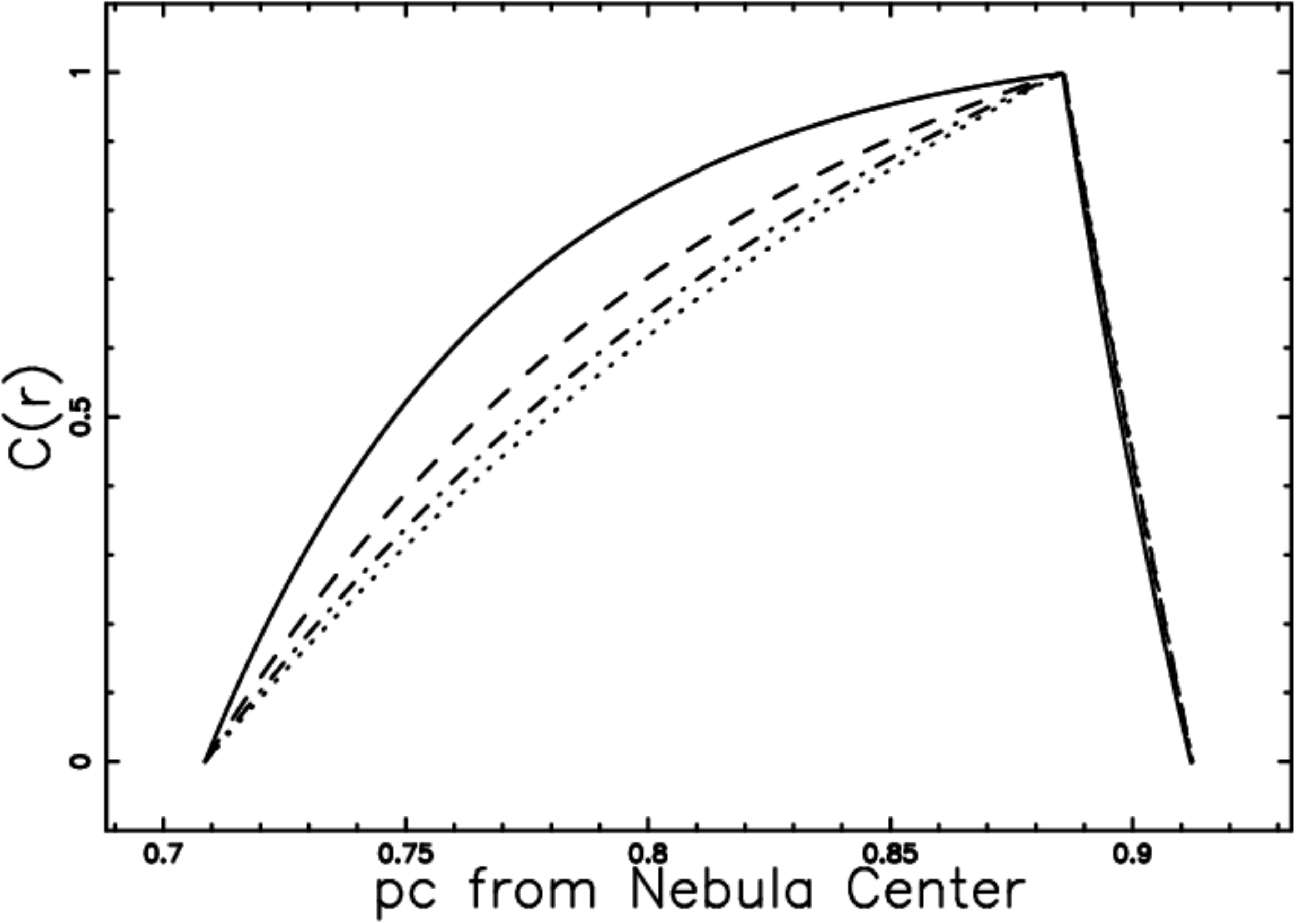}
  \end {center}
\caption {
 Number density of A39  as a
 function of the distance in pc from
 the injection  
 when  $u=1$   , $C_m=1 $, $a=69.6~arcsec $, $b=87~arcsec$ , $c=89.6~arcsec$     
 and   
 $D=2 $       (full line        ),
 $D=7 $       (dashed           ),
 $D=12 $       (dot-dash-dot-dash)
    and
 $D=17 $       (dotted ).
The conversion from $arcsec$ to   $pc$ is done assuming
a distance of 2100 $pc$ for A39.
          }%
    \label{diffusion}
    \end{figure}

The boundary-conditions  give 
\begin{equation}
C_{a,b,drift}(r) =
C_m 
\frac
{ 
e ^{-\frac{u}{D} a} -   e ^{-\frac{u}{D} r} 
}
{
e ^{-\frac{u}{D} a} -   e ^{-\frac{u}{D} b} 
}
\quad a \leq r \leq b ~\quad downstream~side
\quad , 
\label{cab_drift}
\end{equation}
and 
\begin{equation}
C_{b,c,drift}(r) =
C_m 
\frac
{ 
e ^{-\frac{u}{D} c} -   e ^{-\frac{u}{D} r} 
}
{
e ^{-\frac{u}{D} c} -   e ^{-\frac{u}{D} b} 
}
\quad b \leq r \leq c  ~\quad upstream~side
\quad .
\label{cbc_drift}
\end{equation}
A typical plot of the number density for different values
of the diffusion coefficient is reported in 
Figure~\ref{diffusion}.

\subsection{1D diffusion with drift, random walk  }

Given a 1D segment of length $side$ we can implement
the random walk with
step-length  $\lambda$ 
by introducing the numerical parameter $NDIM=\frac{side}{\lambda}$~.
We now report the adopted  rules     
when the  injection is  in the middle of the grid   :  
\begin {enumerate}
\item    The first of the  $NPART$  particles  is chosen.
\item    The random  walk of a particle starts in the middle of
         the grid.
         The probabilities of  having  one step 
         are $p_1$ in the
         negative direction  (downstream)
         ,$ p_1 =  \frac{1}{2} - \mu \times \frac{1}{2}$,
         and $p_2$ in the positive direction  (upstream)
         , $ p_2 =  \frac{1}{2}+ \mu \times \frac{1}{2}$,
         where $\mu$ is a  parameter that characterizes 
         the asymmetry ($0 \leq \mu \leq 1 $).

\item    When the  particle reaches    one of the two 
         absorbing points ,  the motion starts
         another  time from (ii) with
         a different diffusing  pattern.

\item    The number of visits is  recorded on ${\mathcal M}$ ,
         a one--dimensional grid.
\item    The random walk terminates when all  the $NPART$
         particles  are processed.
\item    For the sake  of normalization the
         one--dimensional visitation or number density grid
         ${\mathcal M}$ is divided by  $NPART$.

\end  {enumerate}

\begin{figure*}
\begin{center}
\includegraphics[width=8cm]{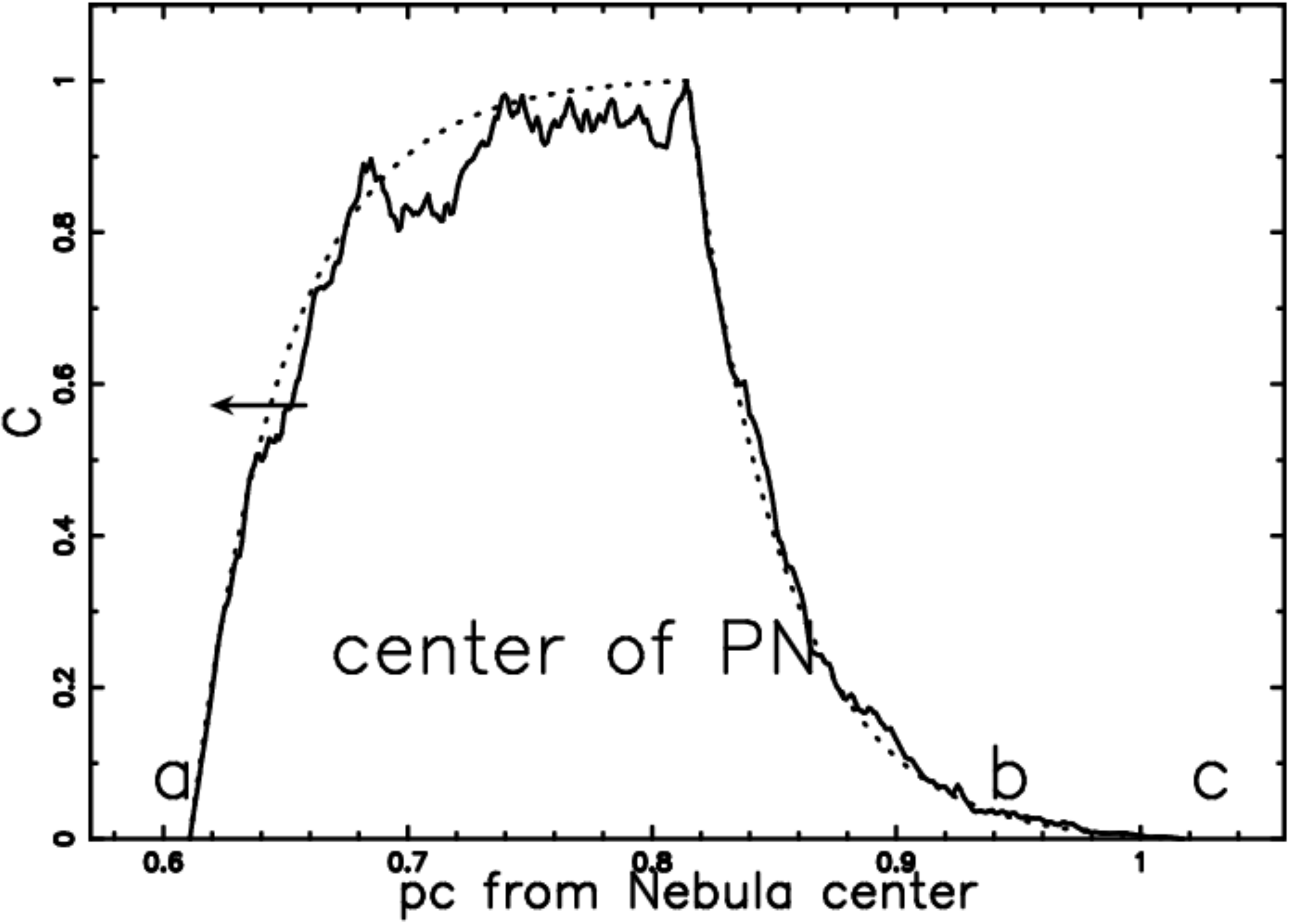}
\end {center}
\caption
{
Number density in A39  of the 1D asymmetric random walk 
(full line), NDIM=401 ,NPART=200 ,$side=40~arcsec$ ,
$\lambda=0.1~arcsec$ 
and  $\mu$ =- 0.013. 
For astrophysical purposes  $\mu$ is negative.
The theoretical number density 
as represented by formulas~(\ref{cab_drift}) and (\ref{cbc_drift})
is  reported     
 when  $u=1$   , $C_m=1 $, $a=60~arcsec$, $b=80~arcsec$ , $c=100~arcsec$     
 and  
 $D=3.84 $       (dotted line ).
The conversion from $arcsec$ to   $pc$ is done assuming
a distance  of 2100 $pc$ for A39.
}
\label{montec}
    \end{figure*}

There is a systematic change of the average particle
position along the
$x$-direction:
\begin{equation}
\langle dx \rangle =    \mu~\lambda \quad,
\label{formula1}
\end {equation}
for each time step. 
If the time step is $dt=\frac{\lambda}{v_{tr}}$
where   ${v_{tr}}$ is the transport velocity,
the asymmetry  ,$\mu$ , 
that characterizes the random walk is  
\begin{equation}
\mu =  \frac{u}{v_{tr}}  
\quad  .
\end{equation}
Figure~\ref{montec} reports ${\mathcal M}(x)$,
the  number  of  visits  
generated by  the Monte Carlo simulation 
as well as  the mathematical solution represented
by formulas~(\ref{cab_drift}) and (\ref{cbc_drift}). 

The solutions of the mathematical   diffusion 
equations~(\ref{cab_drift})   and  (\ref{cbc_drift})
can be rewritten at the light of the 
random walk  and are 
\begin{equation}
C_{a,b,MC}(r) =
C_m 
\frac
{ 
e ^{-\frac{2\mu}{\lambda} a} -   e ^{-\frac{2\mu}{\lambda} r} 
}
{
e ^{-\frac{2\mu}{\lambda} a} -   e ^{-\frac{2\mu}{\lambda} b} 
}
\quad a \leq r \leq b ~\quad downstream~side
\quad , 
\label{cab_MC}
\end{equation}
and 
\begin{equation}
C_{b,c,MC}(r) =
C_m 
\frac
{ 
e ^{-\frac{2\mu}{\lambda} c} -   e ^{- \frac{2\mu}{\lambda} r} 
}
{
e ^{-\frac{2\mu}{\lambda} c} -   e ^{-\frac{2\mu}{\lambda} b} 
}
\quad b \leq r \leq c  ~\quad upstream~side
\quad .
\label{cbc_MC}
\end{equation}

\section{The Image of the PN }
\label{sec_image}
The image of a PN  can be easily modeled once an
analytical or numerical 
law for the intensity of emission as a 
function
of the radial distance from the center is given.
Simple analytical results
for the radial intensity 
 can be deduced in the rim model when 
the length of the layer and the number density are constants
and in the spherical  model when the number density is constant.

The integration of the solutions to the mathematical diffusion
along the line of sight allows us  to deduce 
analytical formulas in the spherical case.
The complexity of the intensity in the aspherical case 
can be attached only from a numerical point of view.

\subsection{Radiative transfer equation}

The transfer equation in the presence of emission only
, see for example  
\cite{rybicki}
 or
\cite{Hjellming1988}
 ,
 is
 \begin{equation}
\frac {dI_{\nu}}{ds} =  -k_{\nu} \zeta I_{\nu}  + j_{\nu} \zeta
\label{equazionetrasfer}
\quad ,
\end {equation}
where  $I_{\nu}$ is the specific intensity , $s$  is the
line of sight , $j_{\nu}$ the emission coefficient,
$k_{\nu}$   a mass absorption coefficient,
$\zeta$ the  mass density at position s
and the index $\nu$ denotes the interested frequency of
emission.
The solution to  equation~(\ref{equazionetrasfer})
 is
\begin{equation}
 I_{\nu} (\tau_{\nu}) =
\frac {j_{\nu}}{k_{\nu}} ( 1 - e ^{-\tau_{\nu}(s)} )
\quad  ,
\label{eqn_transfer}
\end {equation}
where $\tau_{\nu}$ is the optical depth at frequency $\nu$
\begin{equation}
d \tau_{\nu} = k_{\nu} \zeta ds
\quad.
\end {equation}
We now continue analyzing the case of
an
 optically thin layer
in which $\tau_{\nu}$ is very small
( or $k_{\nu}$  very small )
and the density  $\zeta$ is substituted
with our number density C(s) of  particles.
Two cases are taken into account :   the  emissivity is proportional
to the number density and the emissivity is   
proportional to the square of the number density .
In the  linear case 
\begin{equation}
j_{\nu} \zeta =K  C(s)
\quad  ,
\end{equation}
where $K$ is a  constant function.
This can be the case of
synchrotron radiation from an ensemble of 
particles  , see formula (1.175 ) in  \cite{lang} .
This non thermal radiation continuum 
emission was  detected 
in a PN 
associated with a very long-period OH/IR variable star (V1018 Sco),
see~\cite{Cohen2006}.

In the  quadratic  case 
\begin{equation}
j_{\nu} \zeta =K_2  C(s)^2
\quad  ,
\label{eqn_transfer_square}
\end{equation}
where $K_2$ is a  constant function.
This is true for 
\begin{itemize} 
\item 
Free-free radiation from a thermal plasma,
see formula (1.219) in  \cite{lang}  .
This  radiation process was  adopted by 
\cite{Gonzales2006} in the little Homunculus.

\item 
Thermal bremsstrahlung and recombination radiation ,
see formula (1.237) in  \cite{lang}  .
This  radiation process was  adopted 
in PNs by~\cite{Blagrave2006,Schwarz2006,Gruenwald2007}.
\end{itemize}

The intensity is now
\begin{equation}
 I_{\nu} (s) = K
\int_{s_0}^s   C (s\prime) ds\prime \quad  \mbox {optically thin layer}
\quad linear~case \quad ,
\label{transport1}
\end {equation}
or 
\begin{equation}
 I_{\nu} (s) = K_2 
\int_{s_0}^s   C (s\prime)^2 ds\prime \quad  \mbox {optically thin layer}
\quad quadratic~case \quad .
\label{transport2}
\end {equation}
In the Monte Carlo experiments
the number density is memorized  on
the   grid
${\mathcal M}$ and the intensity is
\begin{equation}
{\it I}\/(i,j) = \sum_k  \triangle\,s \times  {\mathcal M}(i,j,k)
\quad  \mbox {optically thin layer}
\quad linear~case 
\quad, 
\label{thin1}
\end{equation}
or 
\begin{equation}
{\it I}\/(i,j) = \sum_k  \triangle\,s \times  {\mathcal M}(i,j,k)^2 
\quad  \mbox {optically thin layer}
\quad quadratic~case 
\quad  ,
\label{thin2}
\end{equation}
where $\triangle$s is the spatial interval between
the various values and  the sum is performed
over the   interval of existence of the index $k$.
The theoretical intensity is then obtained by integrating
the intensity at a given frequency  over the solid angle of the
source.

\subsection{3D Constant Number density in a rim model}

\label{rimsec}
\begin{figure*}
\begin{center}
\includegraphics[width=8cm]{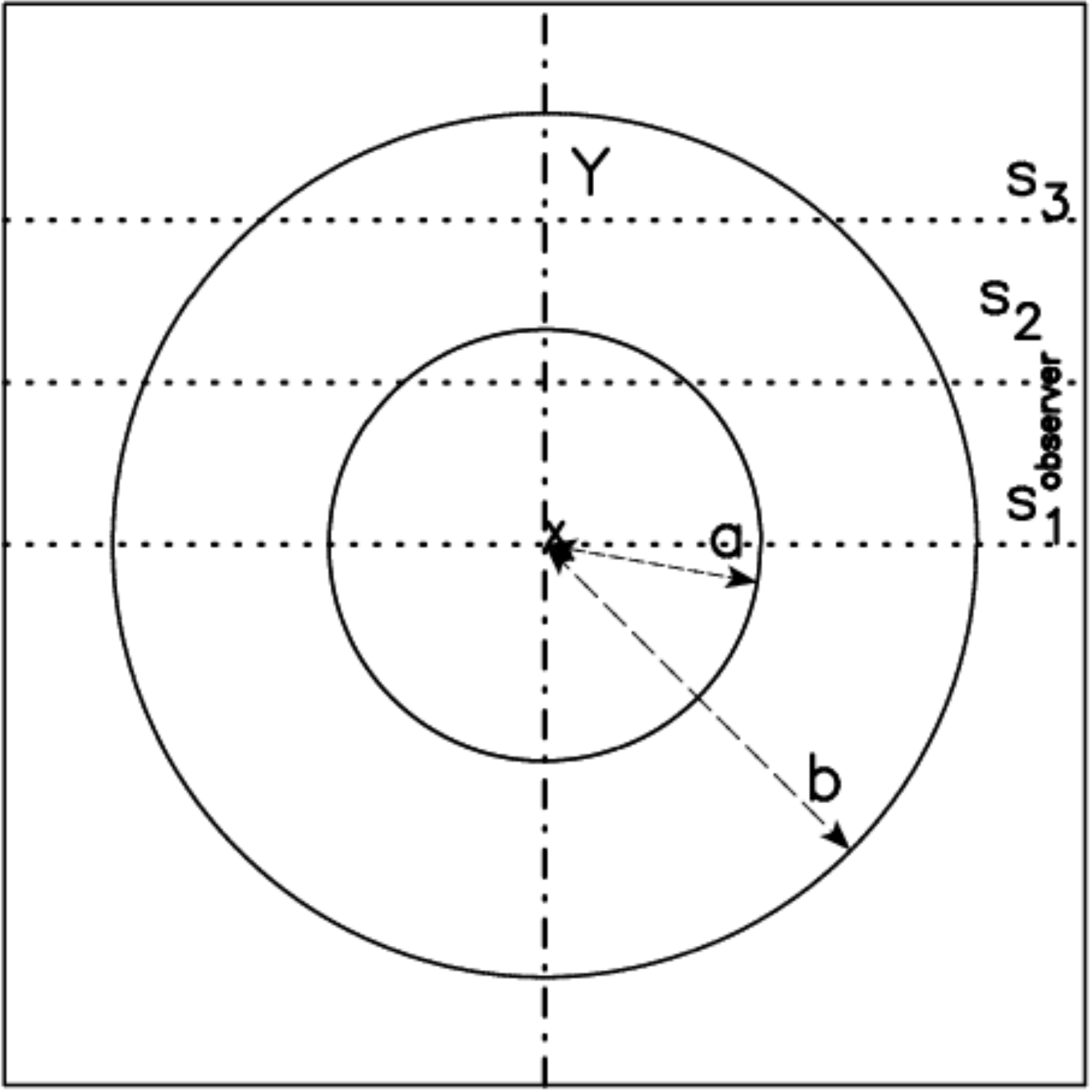}
\end {center}
\caption
{
The two circles (section of spheres)  which   include the region
with constant density
are   represented through
 a full line.
The observer is situated along the x direction, and 
three lines of sight are indicated.
}
\label{plotab}
    \end{figure*}

\begin{figure*}
\begin{center}
\includegraphics[width=8cm]{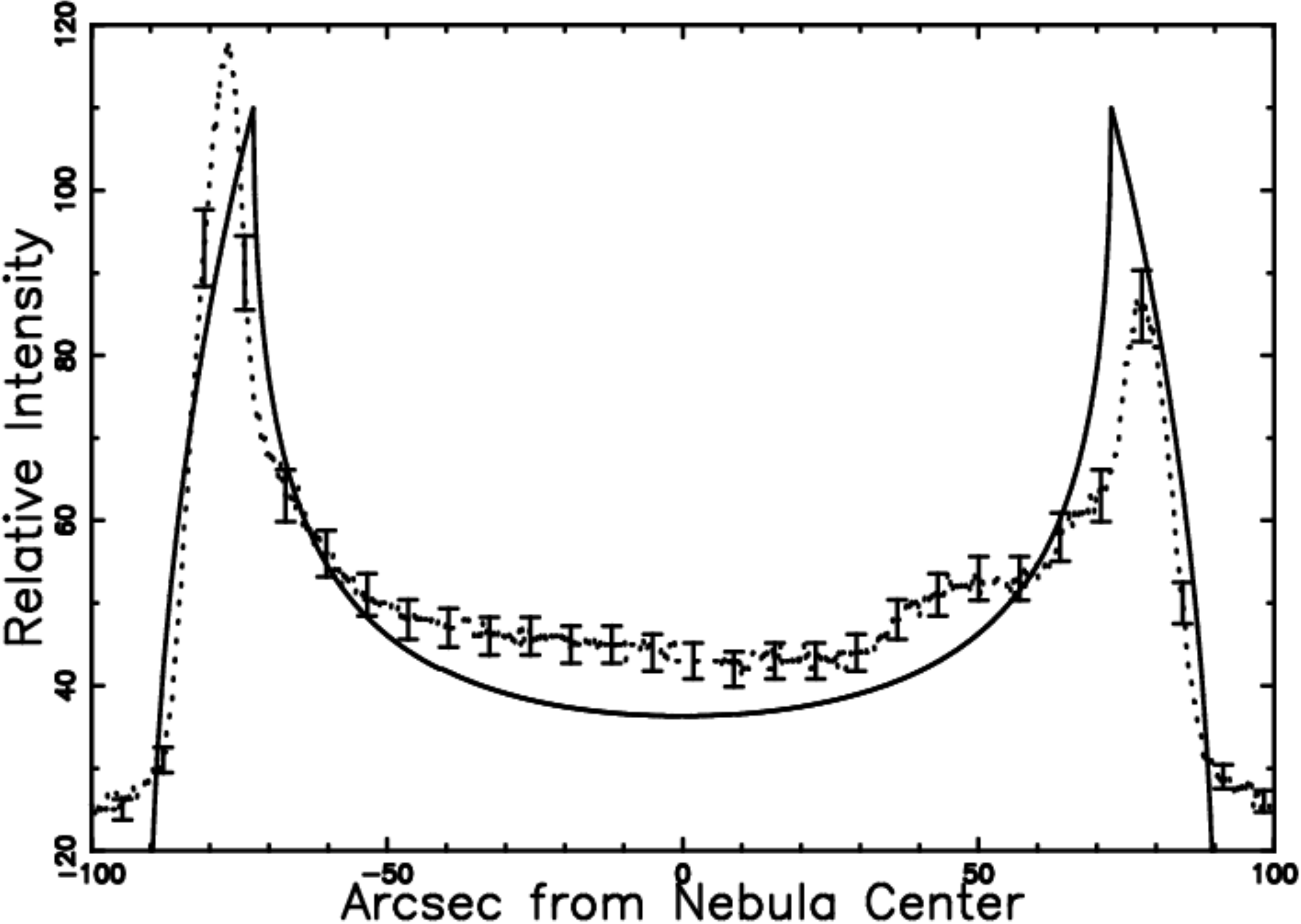}
\end {center}
\caption
{
 Cut of the mathematical  intensity ${\it I}$
 of the rim model ( equation~(\ref{irim})) 
 crossing the center    (full  line  ) of A39 
 and  real data         (dotted line with some error bar ) .
 The number of data is  801  and 
for  this  model $\chi^2$ = 1.487 against $\chi^2$ = 0.862 of
the rim model   
fully described in ~Jacoby et al. (2001).
}
\label{rim_cut_square}
    \end{figure*}

We assume that the number density $C$ is constant and in particular
rises from 0 at $r=a$ to a maximum value $C_m$ , remains
constant up to $r=b$ and then falls again to 0.
This geometrical  description is reported in  
Figure~\ref{plotab}.
The length of sight , when the observer is situated
at the infinity of the $x$-axis , 
is the locus    
parallel to the $x$-axis which  crosses  the position $y$ in a 
Cartesian $x-y$ plane and terminates at the external circle
of radius $b$.
The locus length is   
\begin{eqnarray}
l_{0a} = 2 \times ( \sqrt { b^2 -y^2} - \sqrt {a^2 -y^2}) 
\quad  ;   0 \leq y < a  \nonumber  \\
l_{ab} = 2 \times ( \sqrt { b^2 -y^2})  
 \quad  ;  a \leq y < b    \quad . 
\label{length}
\end{eqnarray}
When the number density $C_m$ is constant between two spheres
of radius $a$ and $b$ 
the intensity of radiation is 
\begin{eqnarray}
I_{0a} =C_m \times 2 \times ( \sqrt { b^2 -y^2} - \sqrt {a^2 -y^2}) 
\quad  ;   0 \leq y < a  \nonumber  \\
I_{ab} =C_m \times  2 \times ( \sqrt { b^2 -y^2})  
 \quad  ;  a \leq y < b    \quad . 
\label{irim}
\end{eqnarray}
The comparison   of observed data of A39 
and the theoretical intensity 
is reported in Figure~\ref{rim_cut_square}
when  data   from  Table~\ref{dataab} are used.

The ratio between the theoretical intensity at the maximum , $(y=b)$ ,
 and at the minimum , ($y=0$) ,  
is given by 
\begin{equation}
\frac {I(y=b)} {I(y=0)} = \frac {\sqrt {b^2 -a^2}} {b-a}
\quad .
\label{ratioteorrim}
\end{equation}
 \begin{table} 
 \caption[]{Simulation of A39 with the rim model } 
 \label{dataab} 
 \[ 
 \begin{array}{llc} 
 \hline 
 \hline 
 \noalign{\smallskip} 
 symbol  & meaning & value  \\ 
 \noalign{\smallskip} 
 \hline 
 \noalign{\smallskip}
a  & radius~of~the~internal~sphere    & 72.5^{\prime\prime} \\ \noalign{\smallskip}
b                 & radius~of~the~external~sphere    & 90.18^{\prime\prime} \\ \noalign{\smallskip}
R_{shell}         & observed~radius~of~the~shell       & 77^{\prime\prime} \\ \noalign{\smallskip}
\delta\,r_{shell,t}  & theoretical~ thickness ~of~the~shell       & 17.6^{\prime\prime} \\ \noalign{\smallskip}
\delta\,r_{shell}  & observed~ thickness ~of~the~shell       & 10.1^{\prime\prime} \\ \noalign{\smallskip}
\frac {I_{limb}} {I_{center}} &  ratio~ of~ observed~ intensities  &
(1.88-2.62)       \\ \noalign{\smallskip} 
\frac {I_{max}} {I(y=0)} & ratio~ of~ theoretical~ intensities  &
3.03      \\ \noalign{\smallskip} 
 \hline 
 \hline 
 \end{array} 
 \] 
 \end {table} 

\subsection{3D Constant Number density in a spherical  model}

\label{spherical}
We assume that the number density $C$ is constant
in a sphere of radius $a$ 
and then falls  to 0.

The length of sight , when the observer is situated
at the infinity of the $x$-axis , 
is the locus    
parallel to the $x$-axis which  crosses  the position $y$ in a 
Cartesian $x-y$ plane and terminates at the external circle
of radius $a$.
The locus  length is   
\begin{eqnarray}
l_{ab} = 2 \times ( \sqrt {a^2 -y^2}) 
\quad  ;   0 \leq y < a  \quad .
\label{lengthsphere}
\end{eqnarray}
When the number density $C_m$ is constant  in the sphere 
of radius $a$ 
the intensity of radiation is 
\begin{eqnarray}
I_{0a} =C_m \times  2 \times ( \sqrt { a^2 -y^2})  
 \quad  ;  0 \leq y < a    \quad . 
\label{isphere}
\end{eqnarray}

\subsection{3D diffusion from a sphere}

\begin{figure*}
\begin{center}
\includegraphics[width=8cm]{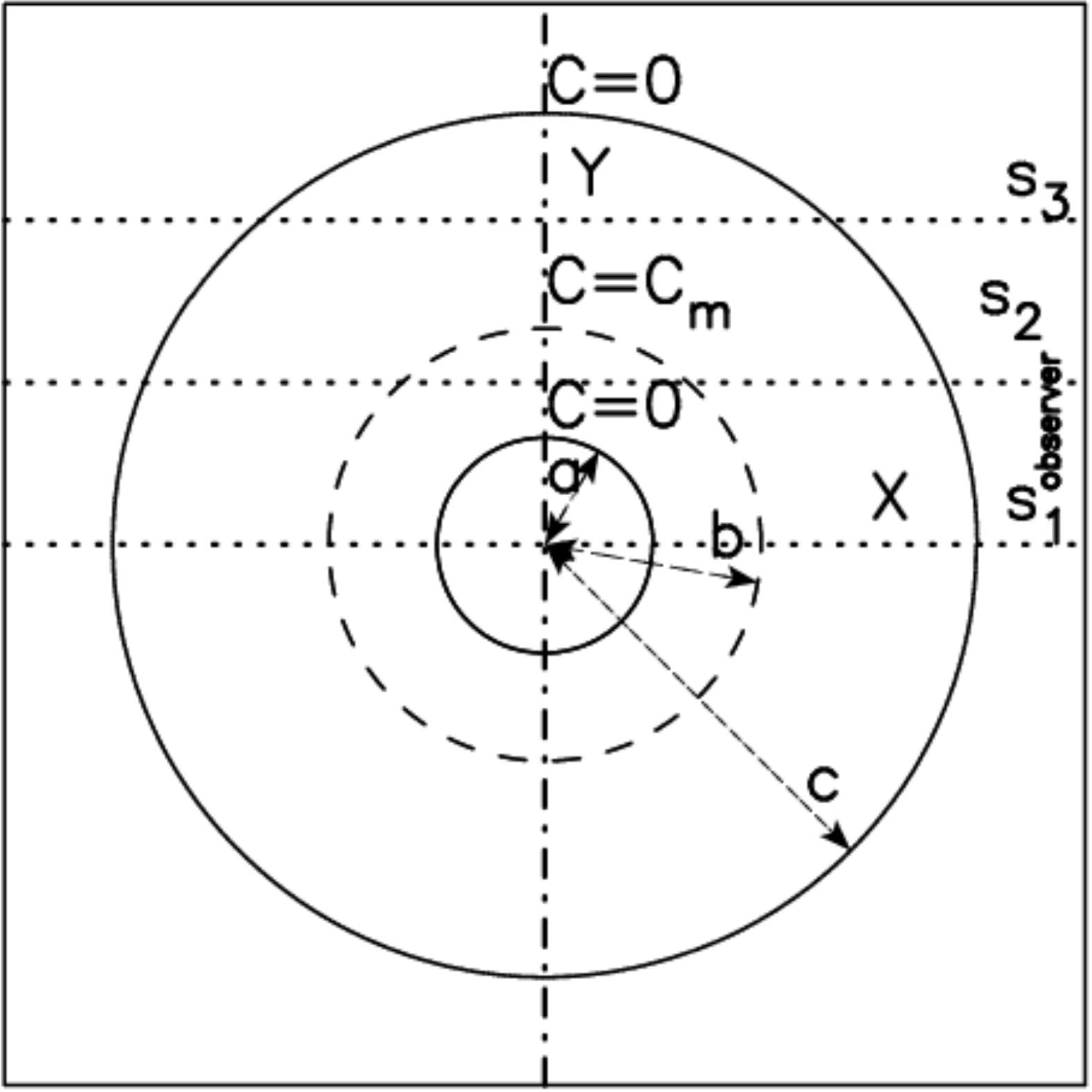}
\end {center}
\caption
{
The spherical source inserted in the great  box is  represented 
through
a dashed line, and the two absorbing boundaries
with a full line.
The observer is situated along the x direction, and 
three lines of sight are indicated.
Adapted from Figure~3.1 by 
Berg (1993) .
}
\label{plot}
    \end{figure*}

Figure~\ref{plot} shows 
a spherical shell  source of radius  $b$
between a spherical absorber
of radius $a$ and a spherical absorber of radius $c$.

The  number density rises from 0 at {\it r=a}  to a
maximum value $C_m$ at {\it r=b} and then  falls again
to 0 at {\it  r=c}~.

The numbers density  to be used 
are formulas (\ref{cab}) and  (\ref{cbc})  once
$r=\sqrt{x^2+y^2}$ is imposed ;
these two numbers density  are  inserted in
formula~(\ref{eqn_transfer_square})  which  
represents the transfer equation with a quadratic dependence
on the number density.
An analogous  case was   solved   in  \cite{Zaninetti2007_c} 
by adopting a linear 
dependence
on  the number density~.
 The     geometry of the phenomena fixes
 three different zones ($0-a,a-b,b-c$) in the variable $y$,
  ;
 the first piece , $I^I(y)$ , is
\begin{eqnarray}
 I^I(y)= \int _{\sqrt{a^2-y^2}} ^{\sqrt{b^2-y^2}} 2 C_{ab}^2 dx
 + \int _{\sqrt{b^2-y^2}} ^{\sqrt{c^2-y^2}}  2C_{bc}^2dx \nonumber\\ 
~= 
-2 
\frac {
{\it C_m}^{2}{b}^{2}
}
{
y ( {b}^{2}-2\,ba+{a}^{2} )  ( {c}^{2}-2\,cb+{b}^{2}
 )
}
\bigl  [ -2\,{a}^{2}\arctan ( {\frac {
\sqrt {{a}^{2}-{y}^{2}}}{y}} ) cb-2\,\sqrt {{a}^{2}-{y}^{2}}ycb
\nonumber \\ 
-
2\,ay\ln  ( \sqrt {{a}^{2}-{y}^{2}}+a ) {b}^{2}+2\,{a}^{2}
\arctan ( {\frac {\sqrt {{b}^{2}-{y}^{2}}}{y}} ) cb
\nonumber \\
+2\,y
\sqrt {{b}^{2}-{y}^{2}}cb+2\,a\ln  ( \sqrt {{b}^{2}-{y}^{2}}+b
 ) y{b}^{2}+2\,a\ln  ( \sqrt {{b}^{2}-{y}^{2}}+b ) y{
c}^{2}
\nonumber  \\
-2\,cy\ln  ( \sqrt {{b}^{2}-{y}^{2}}+b ) {b}^{2}-2\,{
c}^{2}\arctan ( {\frac {\sqrt {{b}^{2}-{y}^{2}}}{y}} ) ba-2
\,y\sqrt {{b}^{2}-{y}^{2}}ba-
\nonumber \\
2\,cy\ln  ( \sqrt {{b}^{2}-{y}^{2}}+
b ) {a}^{2}+2\,c\ln  ( \sqrt {{c}^{2}-{y}^{2}}+c ) y{
b}^{2}
\nonumber  \\
+2\,c\ln  ( \sqrt {{c}^{2}-{y}^{2}}+c ) y{a}^{2}+2\,
\sqrt {{c}^{2}-{y}^{2}}yba+2\,{c}^{2}\arctan ( {\frac {\sqrt {{c}
^{2}
-{y}^{2}}}{y}} ) ba
\nonumber \\
-2\,ay\ln  ( \sqrt {{a}^{2}-{y}^{2}}
+a ) {c}^{2}+\sqrt {{a}^{2}-{y}^{2}}y{c}^{2}
\nonumber  \\
-4\,c\ln  ( 
\sqrt {{c}^{2}-{y}^{2}}+c ) yba+4\,ay\ln  ( \sqrt {{a}^{2}-
{y}^{2}}+a ) cb
\nonumber \\
-\sqrt {{c}^{2}-{y}^{2}}y{a}^{2}+{a}^{2}\arctan
 ( {\frac {\sqrt {{a}^{2}-{y}^{2}}}{y}} ) {b}^{2}-y\sqrt {{
b}^{2}-{y}^{2}}{c}^{2}-{c}^{2}\arctan ( {\frac {\sqrt {{c}^{2}-{y
}^{2}}}{y}} ) {a}^{2}
\nonumber   \\
-\sqrt {{c}^{2}-{y}^{2}}y{b}^{2}+y\sqrt {{b
}^{2}-{y}^{2}}{a}^{2}+{c}^{2}\arctan ( {\frac {\sqrt {{b}^{2}-{y}
^{2}}}{y}} ) {b}^{2}
\nonumber  \\
-{a}^{2}\arctan ( {\frac {\sqrt {{b}^{2
}-{y}^{2}}}{y}} ) {b}^{2}+\sqrt {{a}^{2}-{y}^{2}}y{b}^{2}
\nonumber \\
-{c}^{2
}\arctan ( {\frac {\sqrt {{c}^{2}-{y}^{2}}}{y}} ) {b}^{2}+{
a}^{2}\arctan ( {\frac {\sqrt {{a}^{2}-{y}^{2}}}{y}} ) {c}^
{2} 
\bigr ] 
\\
~ 0 \leq y < a \quad. \nonumber
\label{I_1}
\end{eqnarray}
The second piece , $I^{II}(y)$ , is
 \begin{eqnarray}
 I^{II}(y)=  \int _0 ^{\sqrt{b^2-y^2}} 2 C_{ab}^2 dx
 + \int _{\sqrt{b^2-y^2}} ^{\sqrt{c^2-y^2}}  2C_{bc}^2dx \nonumber\\ 
~= 
2
\frac 
{
\,{b}^{2}{{\it C_m}}^{2} 
}
{
y ( {b}^{2}-2\,ba+{a}^{2} )  ( {c}^{2}-2\,cb+{b}^{2}
 ) 
}
\bigl [ y\sqrt {{b}^{2}-{y}^{2}}{c}^{2}+{a}^{2
}\arctan ( {\frac {\sqrt {{b}^{2}-{y}^{2}}}{y}} ) {b}^{2}
\nonumber \\
-{
c}^{2}\arctan ( {\frac {\sqrt {{b}^{2}-{y}^{2}}}{y}} ) {b}^
{2}
\nonumber  \\
-y\sqrt {{b}^{2}-{y}^{2}}{a}^{2}+{c}^{2}\arctan ( {\frac {
\sqrt {{c}^{2}-{y}^{2}}}{y}} ) {a}^{2}+{c}^{2}\arctan ( {
\frac {\sqrt {{c}^{2}-{y}^{2}}}{y}} ) {b}^{2}+\sqrt {{c}^{2}-{y}
^{2}}y{a}^{2}
\nonumber  \\
+\sqrt {{c}^{2}-{y}^{2}}y{b}^{2}+2\,a\ln  ( y
 ) y{b}^{2}+2\,a\ln  ( y ) y{c}^{2}+2\,cy\ln  ( 
\sqrt {{b}^{2}-{y}^{2}}+b ) {b}^{2}
\nonumber \\
-2\,{a}^{2}\arctan ( {
\frac {\sqrt {{b}^{2}-{y}^{2}}}{y}} ) cb-2\,y\sqrt {{b}^{2}-{y}^
{2}}cb-2\,a\ln  ( \sqrt {{b}^{2}-{y}^{2}}+b ) y{b}^{2}
\nonumber  \\
-2\,a
\ln  ( \sqrt {{b}^{2}-{y}^{2}}+b ) y{c}^{2}-2\,{c}^{2}
\arctan ( {\frac {\sqrt {{c}^{2}-{y}^{2}}}{y}} ) ba-2\,
\sqrt {{c}^{2}-{y}^{2}}yba
\nonumber  \\
-2\,c\ln  ( \sqrt {{c}^{2}-{y}^{2}}+c
 ) y{a}^{2}-2\,c\ln  ( \sqrt {{c}^{2}-{y}^{2}}+c ) y{
b}^{2}+2\,{c}^{2}\arctan ( {\frac {\sqrt {{b}^{2}-{y}^{2}}}{y}}
 ) ba
\nonumber   \\
+2\,y\sqrt {{b}^{2}-{y}^{2}}ba+2\,cy\ln  ( \sqrt {{b}^
{2}-{y}^{2}}+b ) {a}^{2}-4\,a\ln  ( y ) ycb+4\,c\ln 
 ( \sqrt {{c}^{2}-{y}^{2}}+c ) yba
\bigr ]
 \\ 
  a \leq y < b  \quad. 
\nonumber 
\label{I_2}
\end{eqnarray}
The third  piece , $I^{III}(y)$ , is
 \begin{eqnarray}
 I^{III}(y)= \int_0 ^{\sqrt{c^2-y^2}}  2C_{bc}^2 dx \nonumber\\ 
~= 
2
\frac {
{b}^{2}{{\it C_m}}^{2} 
}
{
y \left( {b}^{2}-2\,ba+{a}^{2} \right)  \left( {c}^{2}-2\,cb+{b}^{2}
 \right) 
}
\bigl [ y\sqrt {{c}^{2}-{y}^{2}}{b}^{2}+{c}^{2
}\arctan ( {\frac {\sqrt {{c}^{2}-{y}^{2}}}{y}} ) {b}^{2}
\nonumber \\
-\sqrt {{a}^{2}-{y}^{2}}y{b}^{2}
+{c}^{2}\arctan ( {\frac {\sqrt {{
c}^{2}-{y}^{2}}}{y}} ) {a}^{2}+y\sqrt {{c}^{2}-{y}^{2}}{a}^{2}-
\sqrt {{a}^{2}-{y}^{2}}y{c}^{2}
\nonumber\\
+{a}^{2}\arctan ( {\frac {\sqrt {{
b}^{2}-{y}^{2}}}{y}} ) {b}^{2}
+\sqrt {{b}^{2}-{y}^{2}}y{c}^{2}-
\sqrt {{b}^{2}-{y}^{2}}y{a}^{2}-{c}^{2}\arctan ( {\frac {\sqrt {{
b}^{2}-{y}^{2}}}{y}} ) {b}^{2}
\nonumber  \\
-{a}^{2}\arctan ( {\frac {
\sqrt {{a}^{2}-{y}^{2}}}{y}} ) {b}^{2}
-{a}^{2}\arctan ( {
\frac {\sqrt {{a}^{2}-{y}^{2}}}{y}} ) {c}^{2}+2\,cy\ln  ( 
\sqrt {{b}^{2}-{y}^{2}}+b ) {a}^{2}
\nonumber\\
+2\,ay\ln  ( \sqrt {{a}^
{2}-{y}^{2}}+a ) {c}^{2}
+2\,{a}^{2}\arctan ( {\frac {\sqrt 
{{a}^{2}-{y}^{2}}}{y}} ) cb-2\,a\ln  ( \sqrt {{b}^{2}-{y}^{
2}}+b ) y{c}^{2}
\nonumber \\
-2\,a\ln  ( \sqrt {{b}^{2}-{y}^{2}}+b
 ) y{b}^{2}
-2\,\sqrt {{b}^{2}-{y}^{2}}ycb-2\,{a}^{2}\arctan
 ( {\frac {\sqrt {{b}^{2}-{y}^{2}}}{y}} ) cb
\nonumber \\
+2\,cy\ln 
 ( \sqrt {{b}^{2}-{y}^{2}}+b ) {b}^{2}
-2\,c\ln  ( 
\sqrt {{c}^{2}-{y}^{2}}+c ) y{b}^{2}-2\,c\ln  ( \sqrt {{c}^
{2}-{y}^{2}}+c ) y{a}^{2}
\nonumber \\
-2\,y\sqrt {{c}^{2}-{y}^{2}}ba
-2\,{c}^{
2}\arctan ( {\frac {\sqrt {{c}^{2}-{y}^{2}}}{y}} ) ba+2\,ay
\ln  ( \sqrt {{a}^{2}-{y}^{2}}+a ) {b}^{2}
\nonumber \\
+2\,\sqrt {{a}^{2
}-{y}^{2}}ycb
+2\,\sqrt {{b}^{2}-{y}^{2}}yba+2\,{c}^{2}\arctan ( {
\frac {\sqrt {{b}^{2}-{y}^{2}}}{y}} ) ba
\nonumber\\
+4\,c\ln  ( \sqrt {
{c}^{2}-{y}^{2}}+c ) yba
-4\,ay\ln  ( \sqrt {{a}^{2}-{y}^{2}
}+a ) cb 
\bigr  ] 
\\ 
 b \leq y < c  \quad. 
\label{I_3}
\nonumber
\end{eqnarray}

The profile  of ${\it I}$   made by the three 
pieces (~\ref{I_1}), (~\ref{I_2}) and  (~\ref{I_3}),
can be calibrated on the real data of A39  and an acceptable
match is   realized adopting the parameters 
reported in Table~\ref{dataabc}.
 \begin{table} 
 \caption[]{Simulation of A39 with 3D diffusion } 
 \label{dataabc} 
 \[ 
 \begin{array}{llc} 
 \hline 
 \hline 
 \noalign{\smallskip} 
 symbol  & meaning & value  \\ 
 \noalign{\smallskip} 
 \hline 
 \noalign{\smallskip}
a  & radius~of~the~internal~absorbing~sphere    & 65.96^{\prime\prime} \\ \noalign{\smallskip}
b  & radius~of~the~shock                        & 80^{\prime\prime}  \\ \noalign{\smallskip}
c                 & radius~of~the~external~absorbing~sphere    & 103.5^{\prime\prime} \\ \noalign{\smallskip}
R_{shell}         & observed~radius~of~the~shell       & 77^{\prime\prime} \\ \noalign{\smallskip}
\delta\,r_{shell}  & observed~ thickness ~of~the~shell       & 10.1^{\prime\prime} \\ \noalign{\smallskip}
\frac {I_{limb}} {I_{center}} &  ratio~ of~ observed~ intensities  &
(1.88-2.62)       \\ \noalign{\smallskip} 
\frac {I_{max}} {I(y=0)} & ratio~ of~ theoretical~ intensities  &
2.84              \\ \noalign{\smallskip} 
 \hline 
 \hline 
 \end{array} 
 \] 
 \end {table} 

\begin{figure*}
\begin{center}
\includegraphics[width=8cm]{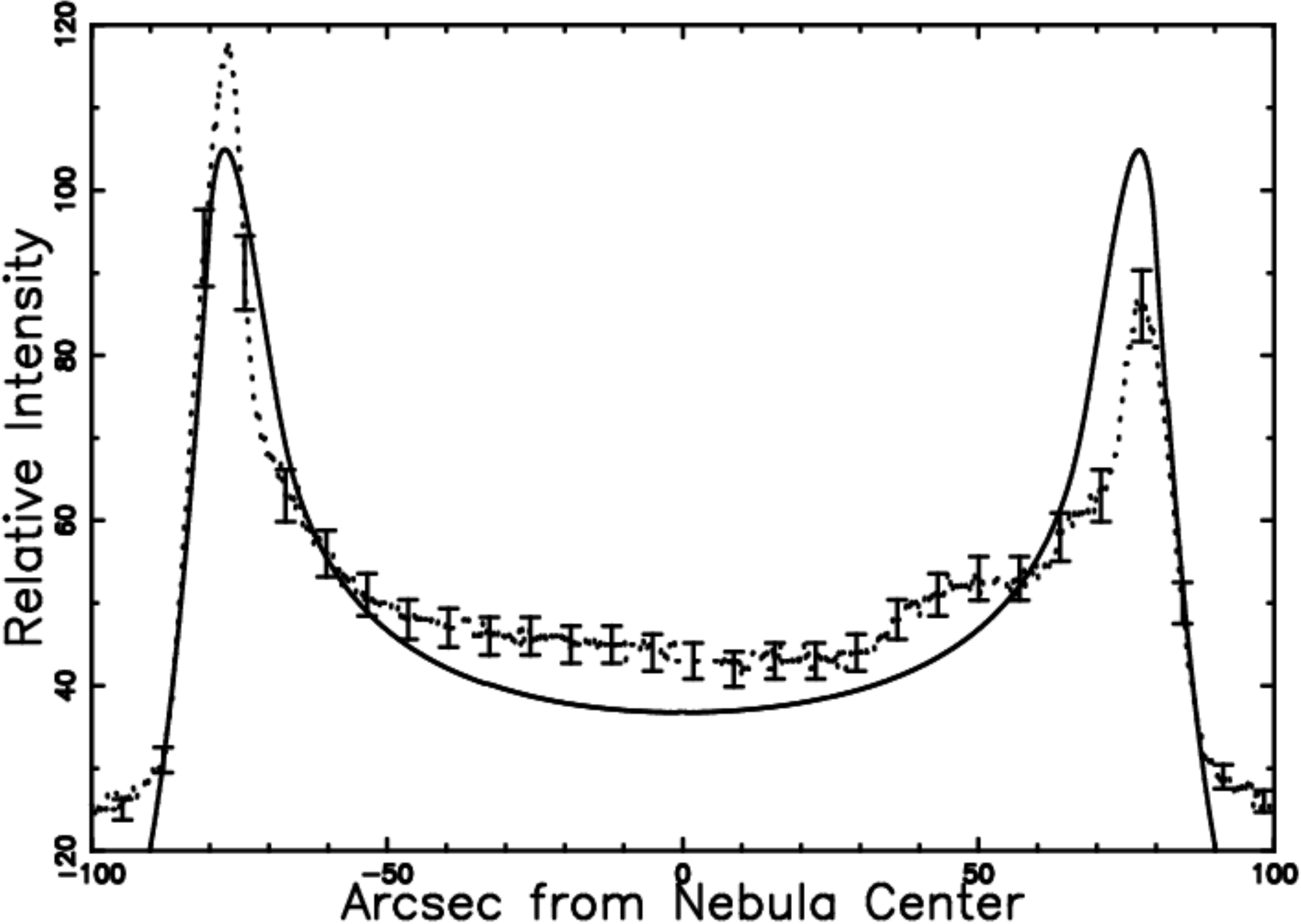}
\end {center}
\caption
{
 Cut of the mathematical  intensity ${\it I}$ 
 (formulas (~\ref{I_1}), (~\ref{I_2}) and  (~\ref{I_3})) ,
 crossing the center    (full    line  ) of A39 
 and  real data         (dotted  line with some error bar).
 The number of data is  801  and 
for  this  model $\chi^2$ = 19.03 against $\chi^2$ = 12.60 of
the rim model   
fully described in ~Jacoby et al. (2001).
}
\label{pn_cut_square}
    \end{figure*}

\begin{figure*}
\begin{center}
\includegraphics[width=8cm]{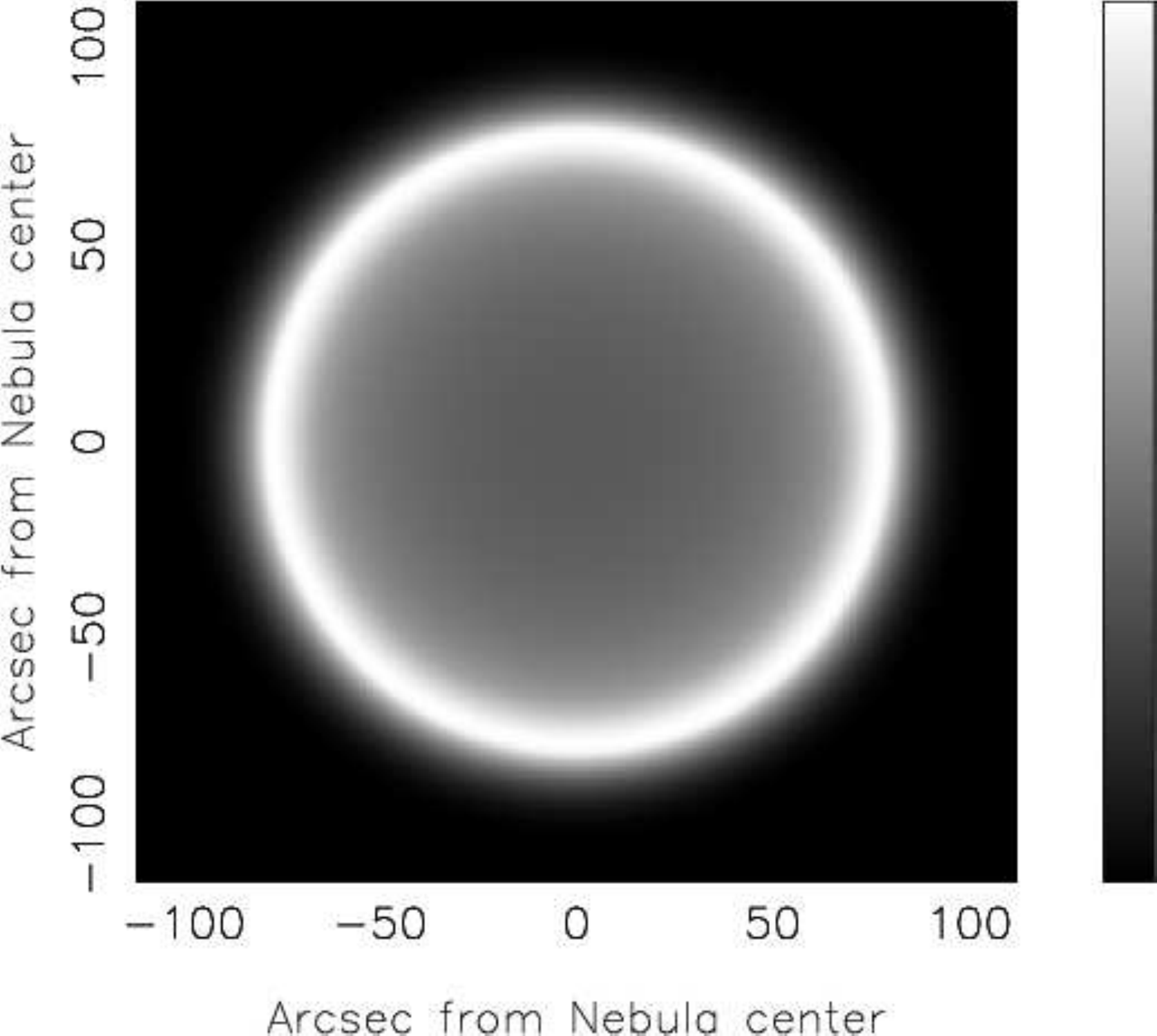}
\end {center}
\caption
{
Contour map  of  ${\it I}$
particularized to simulate  A39.
}
\label{pnbri_square}
    \end{figure*}

The theoretical intensity can therefore 
be plotted as a function of the distance from
the center , see~Figure~\ref{pn_cut_square},
or as  an    image , 
see~Figure~\ref{pnbri_square}.

\begin{figure*}
\begin{center}
\includegraphics[width=8cm]{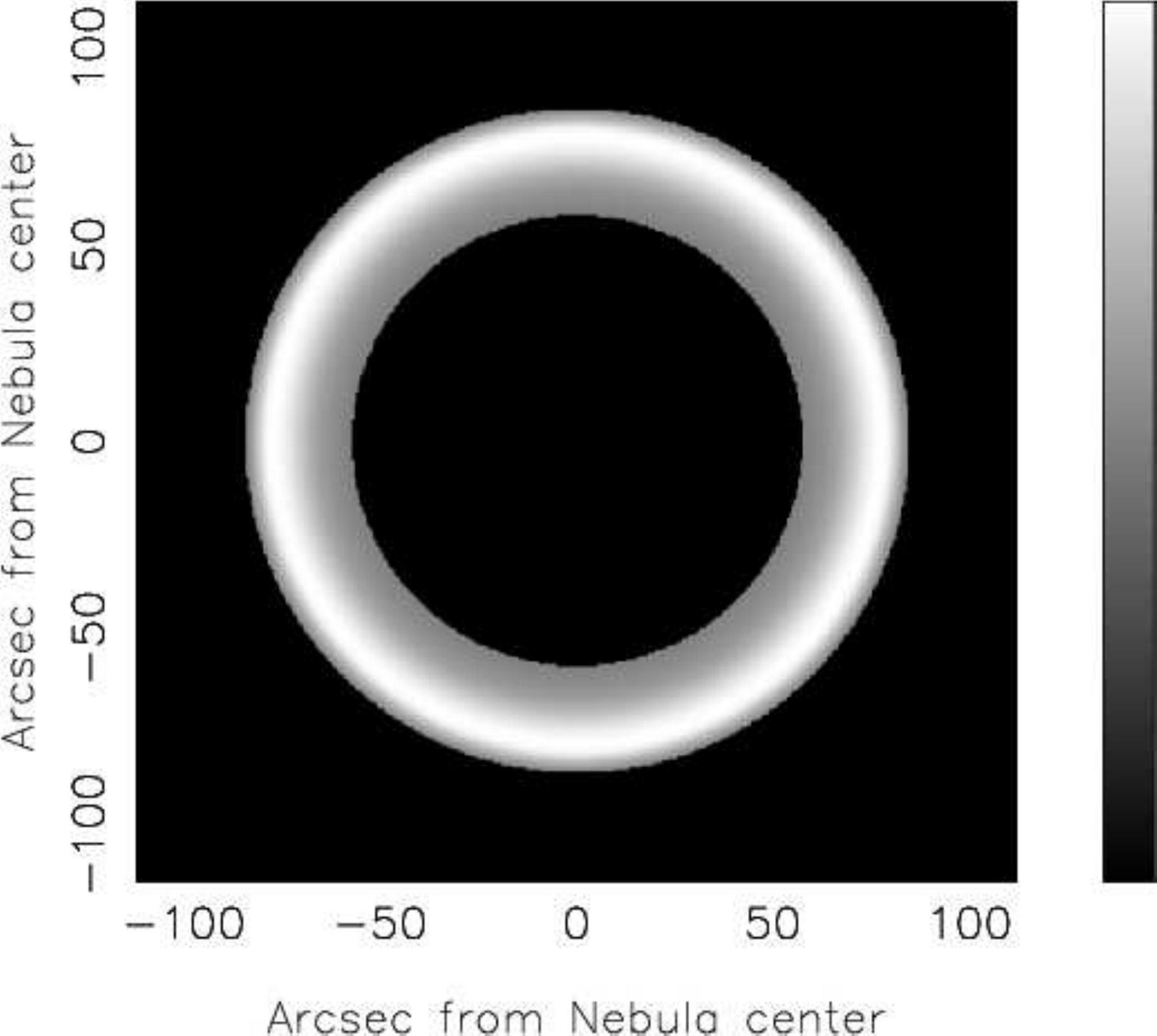}
\end {center}
\caption
{
 The same  as  Figure~\ref{pnbri_square}  but with
 $I_{tr}= I_{max}/2$
}
\label{hole_square}
    \end{figure*}

The effect of the  insertion
of a threshold intensity , $I_{tr}$, given by the
observational techniques , is now analyzed.
The threshold intensity can be parametrized  to  $I_{max}$,
the maximum  value  of intensity characterizing
the ring: a typical  image with  a hole  is visible
in  Figure~\ref{hole_square} when  $I_{tr}= I_{max}/2$.

The position of  the minimum
of ${\it I}$  is at $y=0$ and the position of the maximum
is situated at $y=b$.

The ratio between the theoretical intensity at maximum 
, $I_{max}$ at $y=b$  ,
 and at the minimum ($y=0$) 
is given by 
\begin{equation}
\frac {I_{max}} {I(y=0)} = \frac{Numerator}{Denominator}
\quad,
\label{ratioteor}
\end {equation}
where 
\begin{eqnarray}
 Numerator=  
( {b}^{2}-2\,ba+{a}^{2} ) 
\nonumber\\
\times  ( 2\,cb\ln  ( b
 ) -2\,c\ln  ( \sqrt {{c}^{2}-{b}^{2}}+c ) b+b\sqrt {
{c}^{2}-{b}^{2}}+{c}^{2}\arctan ( {\frac {\sqrt {{c}^{2}-{b}^{2}}
}{b}} )  ) 
 \quad,
\end{eqnarray}
and 
\begin{eqnarray}
 Denominator= \nonumber\\ 
2\,b ( {a}^{2}c-{c}^{2}a-2\,bca\ln  ( a ) +2\,bca\ln 
 ( c ) -b{a}^{2}+b{c}^{2}-{b}^{2}c+{b}^{2}a+{b}^{2}a\ln 
 ( a ) 
\nonumber \\
-{c}^{2}a\ln  ( b ) +{b}^{2}c\ln  ( 
b ) -{b}^{2}a\ln  ( b ) -{b}^{2}c\ln  ( c
 ) +{a}^{2}c\ln  ( b ) +{c}^{2}a\ln  ( a
 ) -{a}^{2}c\ln  ( c )  ) 
\quad.
\end{eqnarray}

The  ratio rim(maximum) /center(minimum)  of the observed 
intensities  as well 
as 
the theoretical one  are 
reported in Table~\ref{dataabc}.

\begin{figure*}
\begin{center}
\includegraphics[width=8cm]{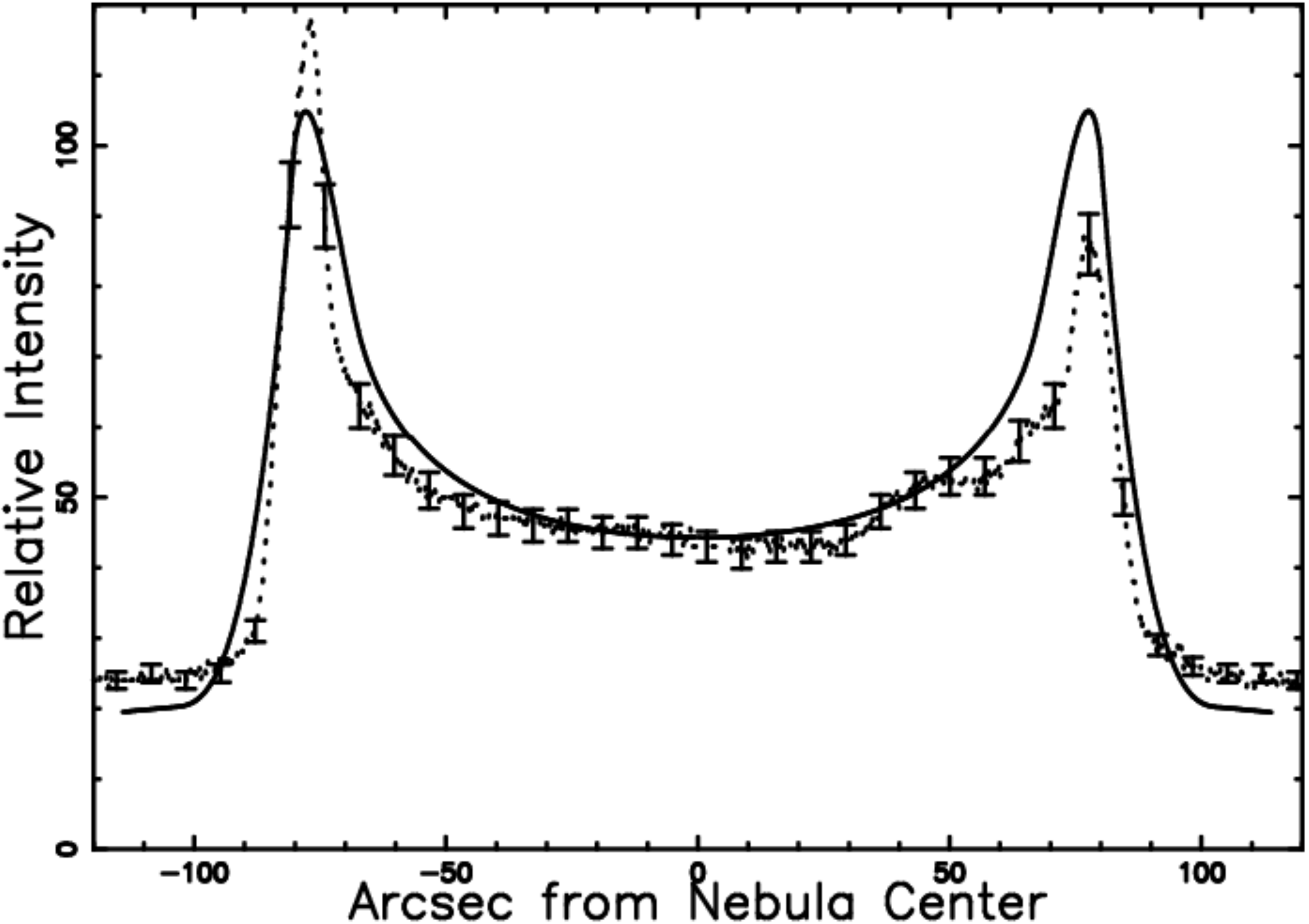}
\end {center}
\caption
{
 Cut of the mathematical  intensity ${\it I}$
 that characterizes the two-phase diffusion 
 ( full  line  )  
 and  real data   of A39       (dotted  line with some error bar).
 The number of data is  801  and 
for  this  model $\chi^2$ = 2.29 against $\chi^2$ = 12.606 of
the rim model   
fully described in ~Jacoby et al. (2001).
}
\label{pn_halo_square}
    \end{figure*}

\begin{figure*}
\begin{center}
\includegraphics[width=8cm]{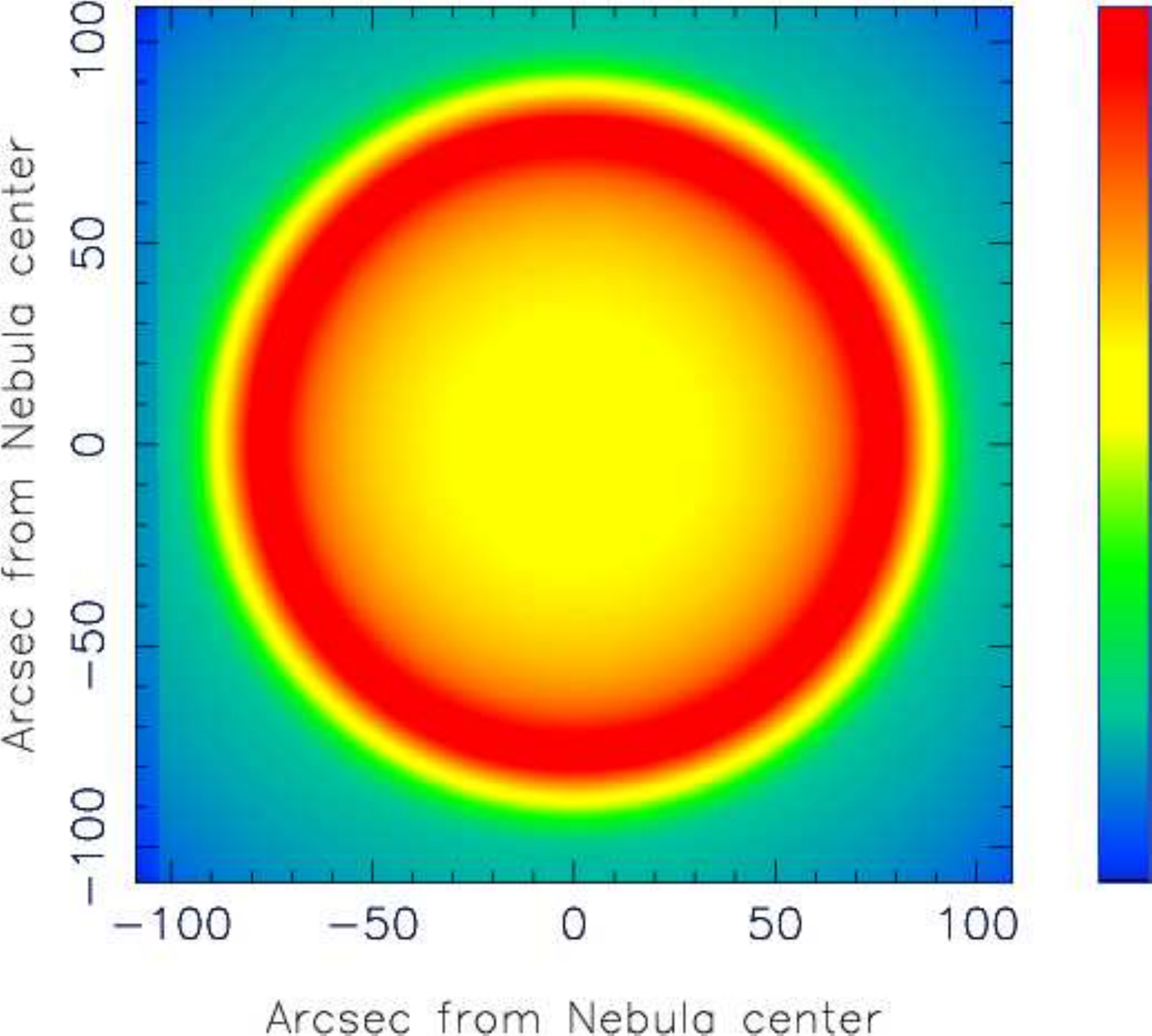}
\end {center}
\caption
{
Contour map of the decimal logarithm   of  ${\it I}$
of the   two-phase  diffusion 
relative to  A39.
}
\label{pnbrihalo_square_colorlog}
    \end{figure*}

Up to now we have not  described   the fainter halo of A39 
which 
according to \cite{Jacoby2001}  extends $15^{\prime\prime}$ 
beyond the rim.
The halo intensity can be modeled by introducing 
two different processes of diffusion characterized 
by different geometrical  situations .
The first is represented 
by ${\it I}$   made by the three 
pieces (~\ref{I_1}), (~\ref{I_2}) and  (~\ref{I_3}),
the second one is the intensity   between a  larger sphere
( $r=2 \times c$) 
and  smaller sphere  ( $r=b$) 
with constant density , see formula~(\ref{irim} )
\begin{equation}
I = I(C_{m,1},a_1,b_1,c_1,y) + I_{0a} (C_{m,2},a_2,b_2,y)
\quad  ,
\label{isum}
\end{equation}
where the  numbers 1 and 2  stand for first process
and second process .
The second process   with  constant density  will be characterized 
by a larger volume of the considered  bigger sphere and 
smaller number density , i.e.  $C_{m,2} \ll C_{m,1}$  .
A typical  result of this two phase  process  is plotted  
in Figure~\ref{pn_halo_square} 
and the image reported in Figure~\ref{pnbrihalo_square_colorlog} ;
the adopted 
parameters  are reported in Table~\ref{dataabchalo}.

 \begin{table} 
 \caption[]{Simulation of A39 , halo comprised  } 
 \label{dataabchalo} 
 \[ 
 \begin{array}{llc} 
 \hline 
 \hline 
 \noalign{\smallskip} 
 symbol  & meaning & value  \\ 
 \noalign{\smallskip} 
 \hline 
 \noalign{\smallskip}
a_1  & radius~of~the~internal~absorbing~sphere    & 66.3^{\prime\prime} \\ \noalign{\smallskip}
b_1  & radius~of~the~shock                        & 80^{\prime\prime}  \\ \noalign{\smallskip}
c_1  & radius~of~the~external~absorbing~sphere    & 103.5^{\prime\prime} \\ \noalign{\smallskip}
C_{m,1} &  maximum~number density~main~diffusion         & 1                 \\ \noalign{\smallskip}
a_2  & internal~radius~of~the~halo~process        & 80 ^{\prime\prime} \\ \noalign{\smallskip}
b_2  & external~radius~of~the~halo~process        & 207 ^{\prime\prime} \\ \noalign{\smallskip}
C_{m,2} &number density~halo                       & 0.045                \\ \noalign{\smallskip}
\frac {I_{limb}} {I_{center}} &  ratio~ of~ simulated ~ intensities  &
2.85      \\ \noalign{\smallskip} 
 \hline 
 \hline 
 \end{array} 
 \] 
 \end {table}

\subsection{3D diffusion from a sphere with drift}
\label{secdrift}
\begin{figure*}
\begin{center}
\includegraphics[width=8cm]{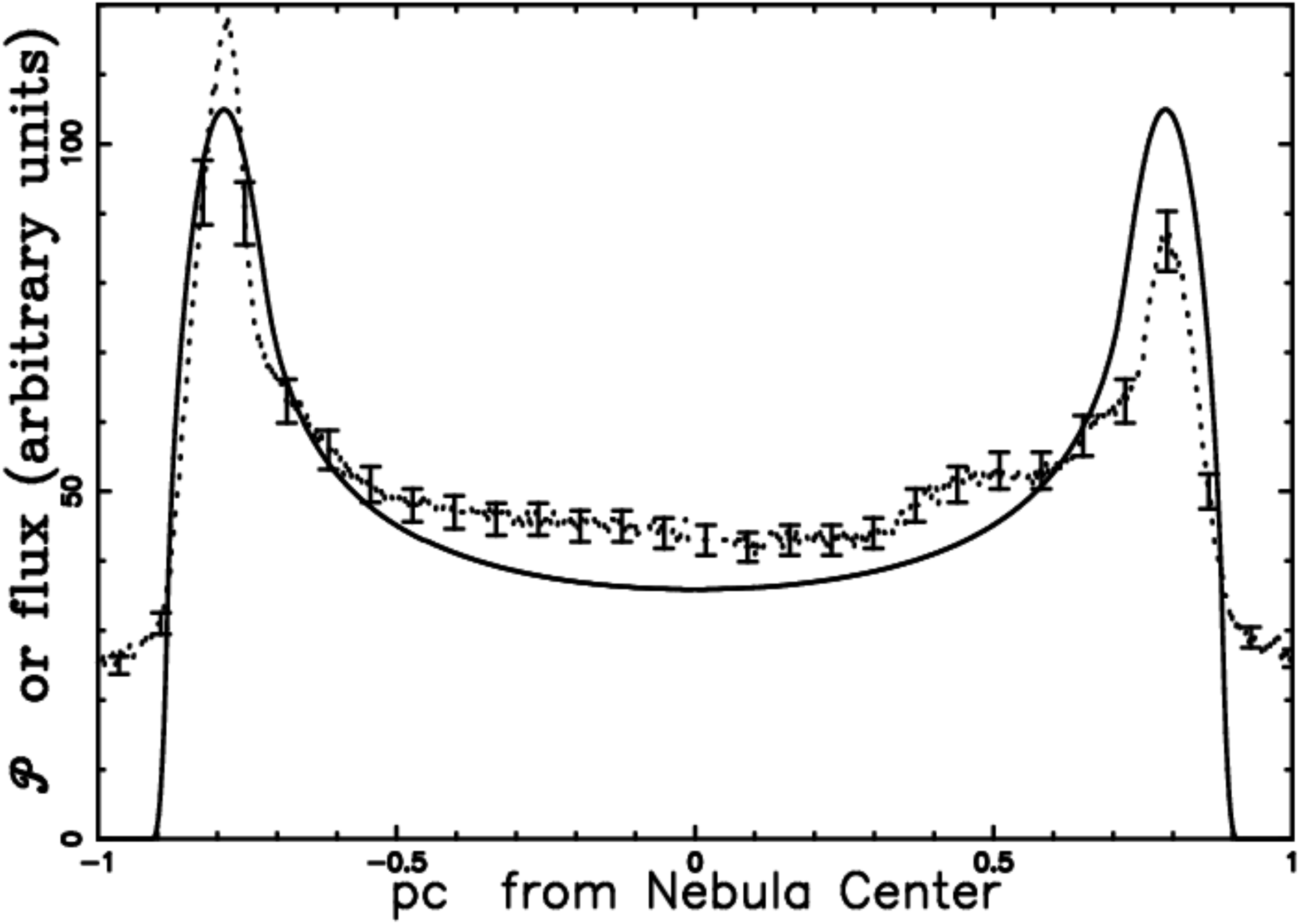}
\end {center}
\caption
{
 Cut of the numerical  intensity ${\it I}$ 
 crossing the center    (dotted  line  ) of A39 
 when the  drift is considered 
 and  real data         (full line).
 The parameters are 
  $u=1$   , $a=69.6~arcsec $, $b=87~arcsec  $ , $c=89~arcsec  $     
 and   
 $D=6 $.
 In this case $\frac {I_{max}} {I(y=0)}$=2.92 .
 The number of data is  801  and 
for  this  model $\chi^2$ = 20.96 against $\chi^2$ = 10.36 of
the rim model   
fully described in ~Jacoby et al. (2001).
The conversion from $arcsec$ to   $pc$ is done assuming
a distance of 2100 $pc$ for A39.
}
\label{pn_cut_asym_2}
    \end{figure*}

The influence of advection on diffusion can be explored 
assuming that in 3D  the number density scales 
in the radial direction 
as does  the 1D solution with drift 
represented by 
formulas (\ref{cab_drift}) and  (\ref{cbc_drift})  . This 
is an approximation
due  to the absence of Fick's second equation
in 3D.
Also here  the     geometry of the phenomena fixes
three different zones ($0-a,a-b,b-c$) in the variable $y$ , see 
Figure~\ref{plot},  and 
the intensity along the line of sight can be found by imposing
$r=\sqrt{x^2+y^2}$ .
In this case,  the integral operation 
of the square of the number density 
which  gives the intensity 
can be performed only numerically  ,
see  Figure~\ref{pn_cut_asym_2}.

\subsection{3D complex morphologies }

\label{seccomplex}
\begin{figure}
  \begin{center}
\includegraphics[width=8cm]{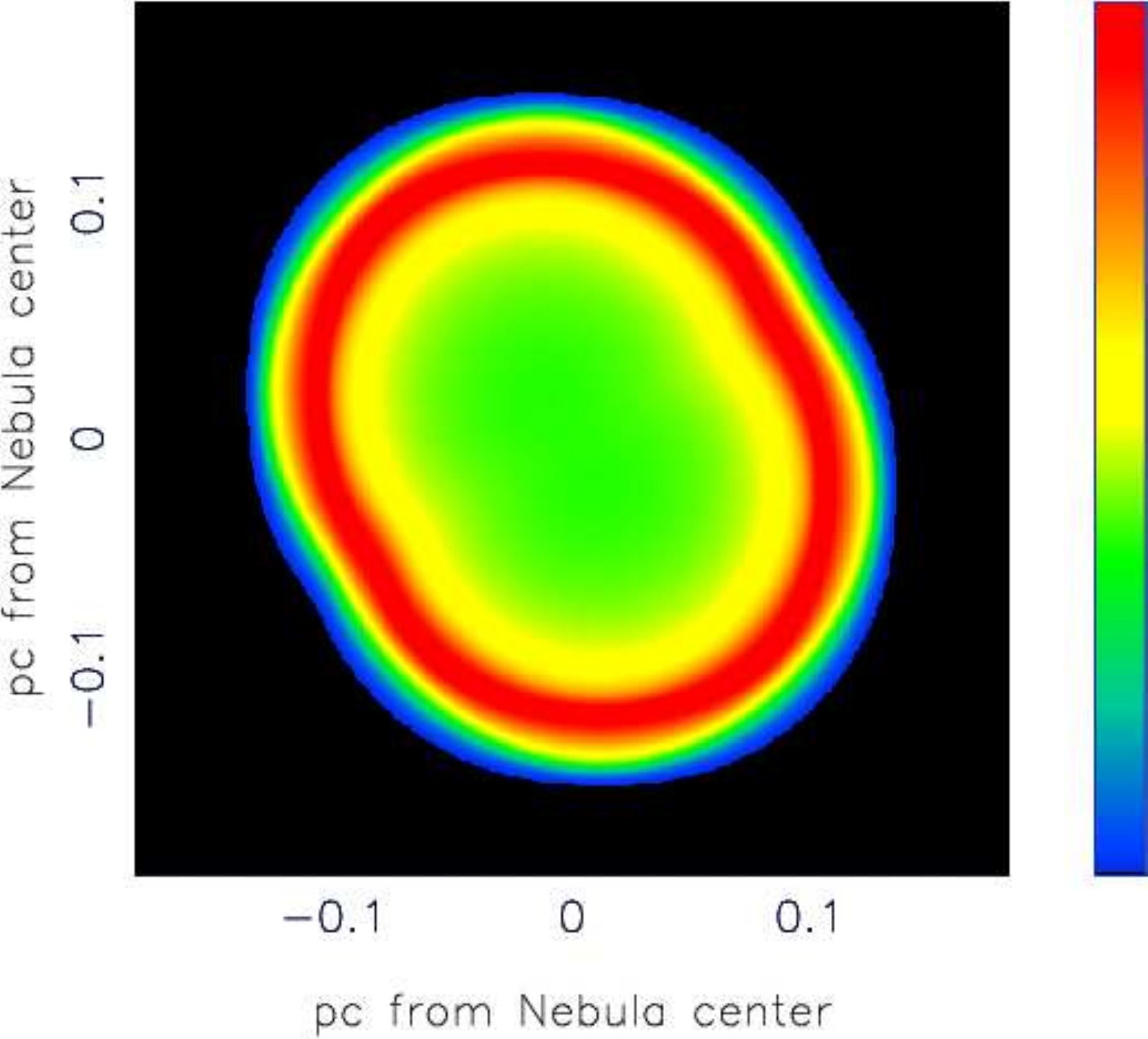}
  \end {center}
\caption {
Map of the theoretical intensity  of the Ring nebula.
Physical parameters as in Table~\ref{parameters} and
$f$=12 .
 The three Eulerian angles 
 characterizing the point of view are 
     $ \Phi $=180    $^{\circ }  $, 
     $ \Theta $=90   $^{\circ }$
and  $ \Psi $=-30    $^{\circ }   $.
          }%
    \label{ring_heat}
    \end{figure}

\begin{figure*}
\begin{center}
\includegraphics[width=8cm]{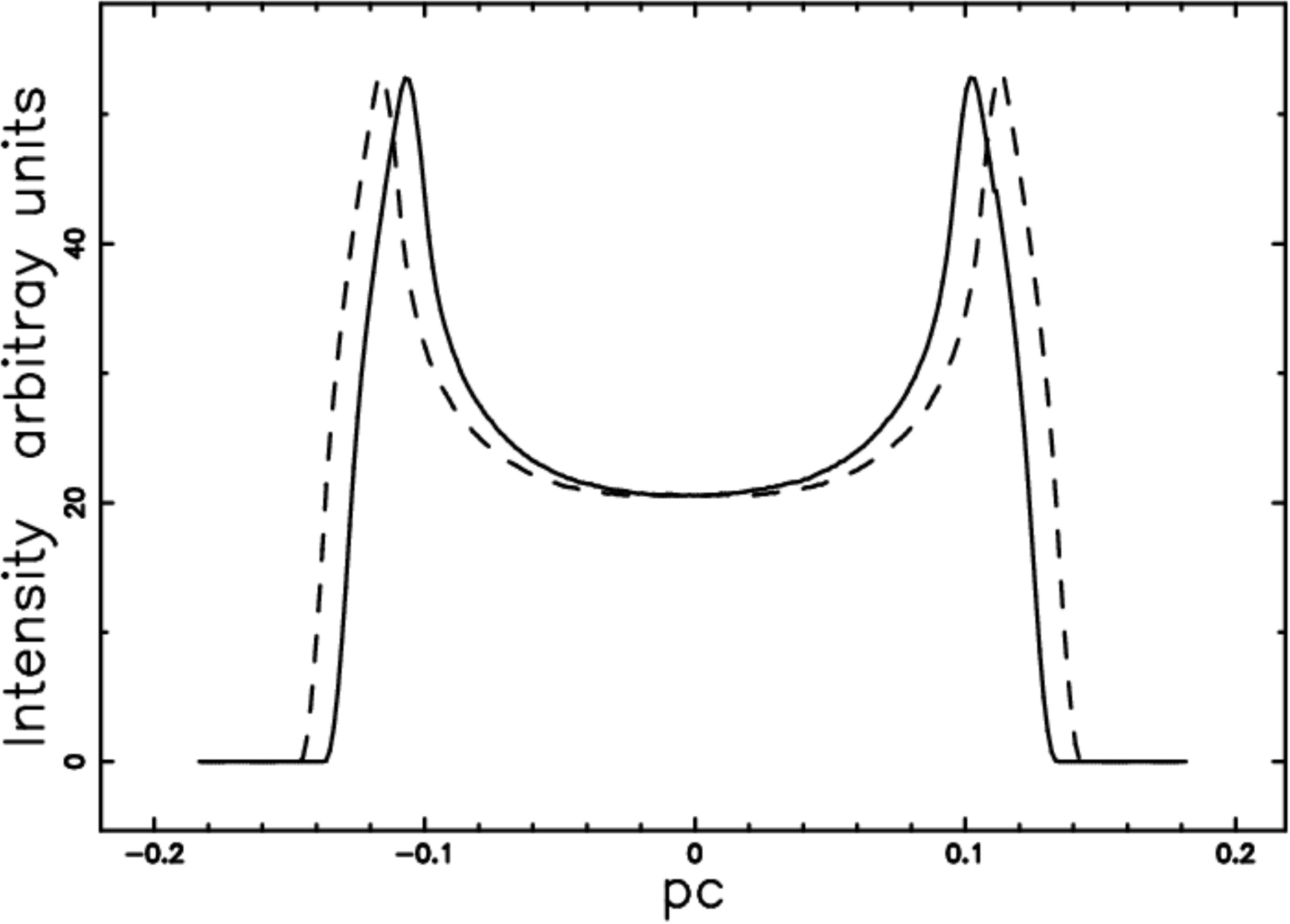}
\end {center}
\caption
{
 Two  cut of the mathematical  intensity ${\it I}$
 crossing the center    of the Ring nebula:
 equatorial cut          (full line)
 and  polar cut          (dotted line) .
}
\label{cut_xy_ring}
    \end{figure*}

The numerical approach to the intensity map can be implemented
when the ellipsoid that characterizes the expansion surface 
of the PN has a constant thickness  expressed , for example ,
as  $r_{min}/f $ where $r_{min}$ is the minimum radius of the
ellipsoid and $f$ an integer.
We remember that  $f=12$ has a physical basis 
in the symmetrical case  , see~\cite{Dalgarno1987}.
The numerical algorithm that allows us  to build the 
image  is now outlined 
\begin{itemize}
\item
A memory grid  ${\mathcal {M}} (i,j,k)$ that  contains 
$NDIM^3$ pixels is considered
\item
The points of the thick ellipsoid  are memorized
on  ${\mathcal {M}}$
\item
Each point of  ${\mathcal {M}} $  has spatial coordinates
$x,y,z$ 
which  can be  represented  by
the following $1 \times 3$  matrix ,$A$,
\begin{equation}
A=
 \left[ \begin {array}{c} x \\\noalign{\medskip}y\\\noalign{\medskip}{
\it z}\end {array} \right] 
\quad  .
\end{equation}
The point of view of the observer is characterized by the
Eulerian  angles   $(\Phi, \Theta, \Psi)$
and  therefore  by a total rotation 
 $3 \times 3$  matrix ,
$E$ , see \cite{Goldstein2002}. 
The matrix point  is now 
represented by the following $1 \times 3$  matrix , $B$,
\begin{equation}
B = E \cdot A 
\quad .
\end{equation}
\item 
The map in intensity is obtained by summing the points of the 
rotated images along a direction 
, for example along z , 
 ( sum over the range of one index, for example k ).
\end{itemize}
Figure~\ref{ring_heat} reports the rotated image 
of the Ring nebula and  Figure~\ref{cut_xy_ring} 
reports two cuts along the polar and equatorial 
directions.

\begin{figure*}
\begin{center}
\includegraphics[width=8cm]{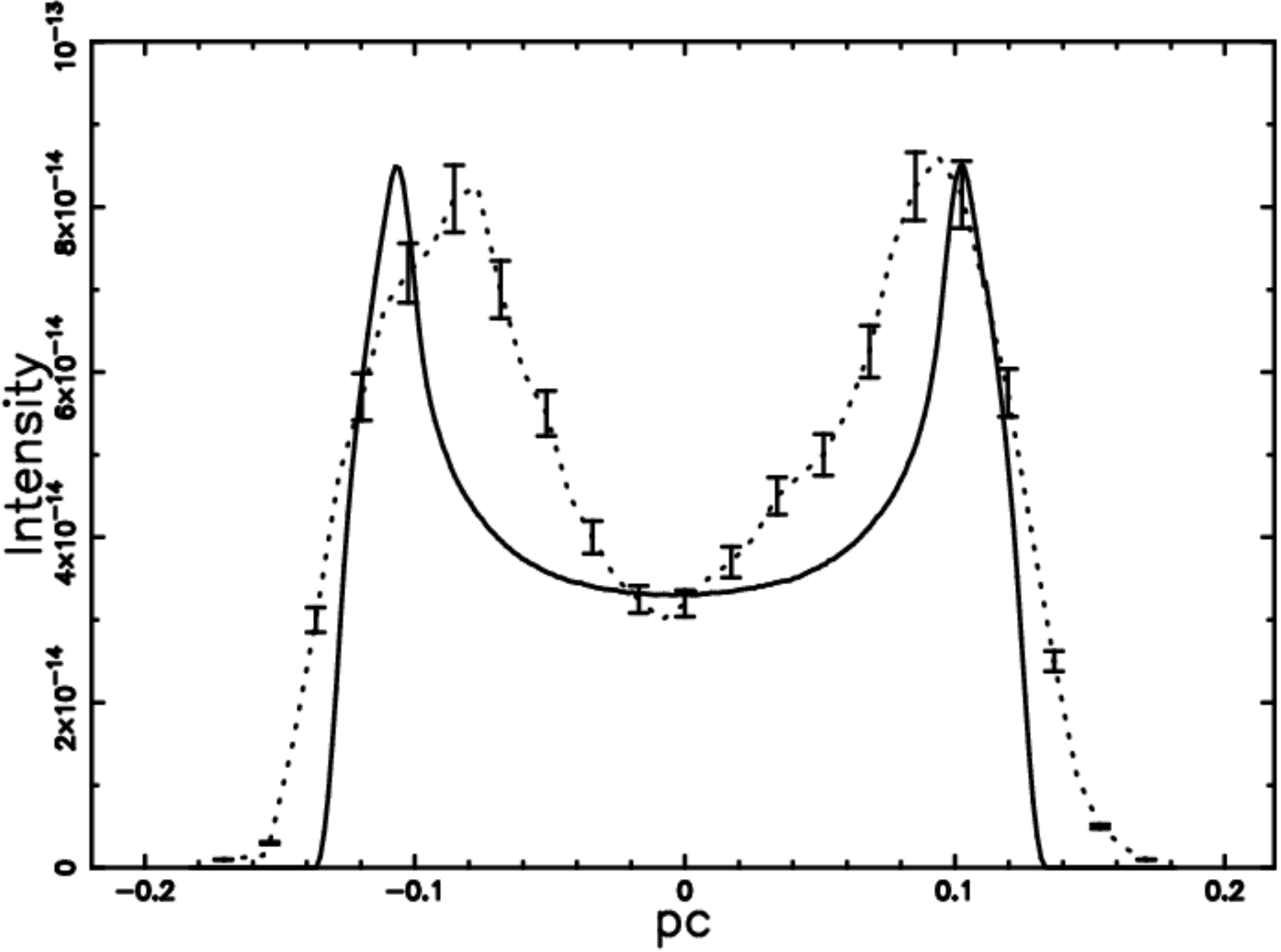}
\end {center}
\caption
{
 Cut of the mathematical  intensity ${\it I}$
 of the Ring Nebula      
 crossing the center    (full  line  ) 
 and  real data  of $H_{\beta}$ 
 (dotted line with some error bar ) .
 The number of data is  250  and 
for  this  model $\chi^2$ = 15.53~.
The real data are extracted  by the author 
from Figure 1 of Garnett and Dinerstein 2001.
}
\label{cut_confronto}
    \end{figure*}

Figure~\ref{cut_confronto} reports 
the comparison between  a theoretical and observed
east-west cut in $H_{\beta}$ 
that cross the center of the nebula, see Figure~1 in 
\cite{Garnett2001}.

A comparison can be made   with the color composite image
of Doppler-shifted $H_2$ emission as represented
in Figure~2 in \cite {Hiriart2004}.

\begin{figure}
  \begin{center}
\includegraphics[width=8cm]{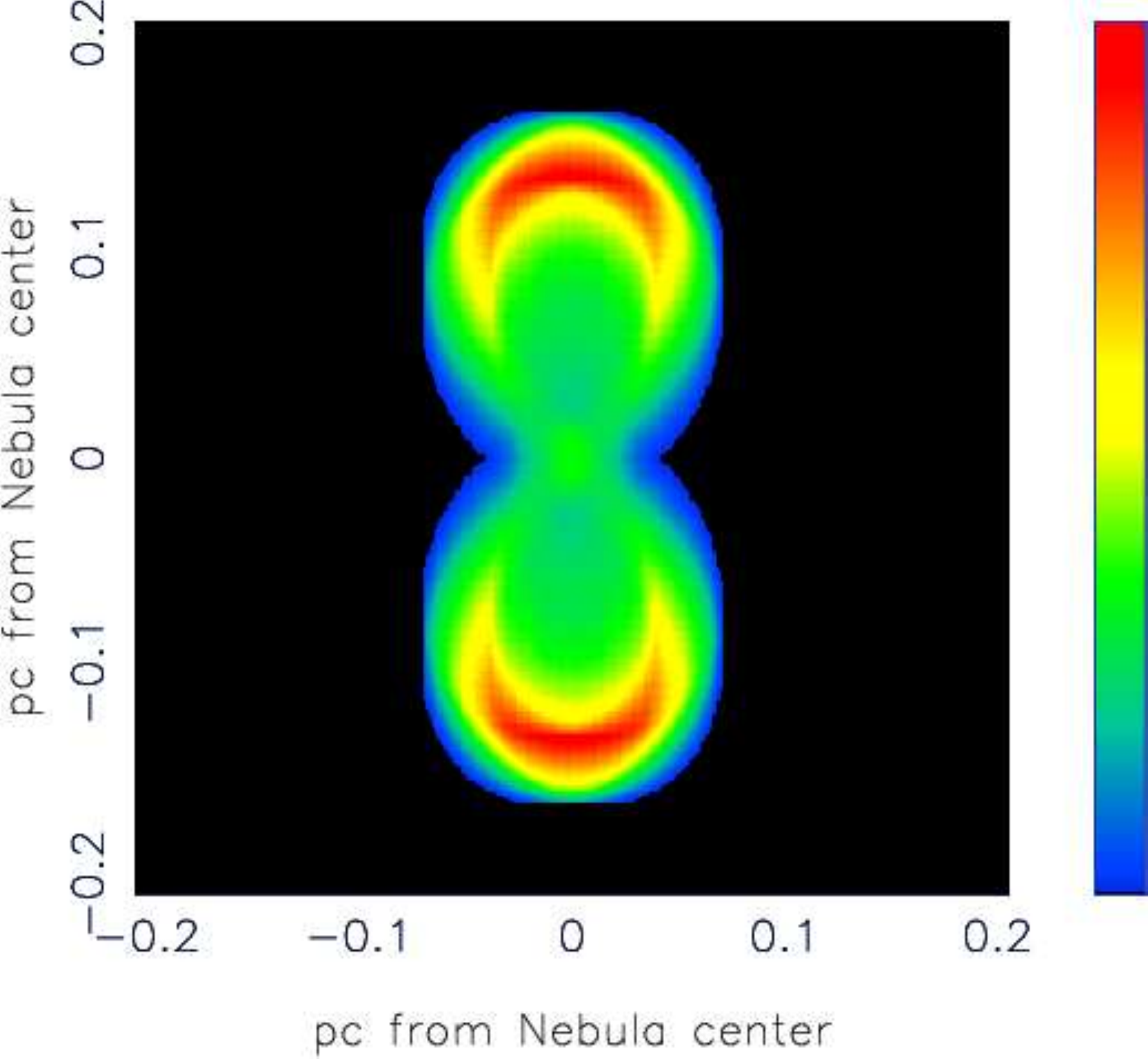}
  \end {center}
\caption {
Map of the theoretical intensity  of  MyCn 18 .
Physical parameters as in Table~\ref{parametershom} and
$f$=12 .
The three Eulerian angles 
characterizing the point of view are $ \Phi $=180   $^{\circ }$
, $ \Theta $=90   $^{\circ }$
and  $ \Psi $=0  $^{\circ }$.
          }%
    \label{mycn18_heat}
    \end{figure}

\begin{figure}
  \begin{center}
\includegraphics[width=8cm]{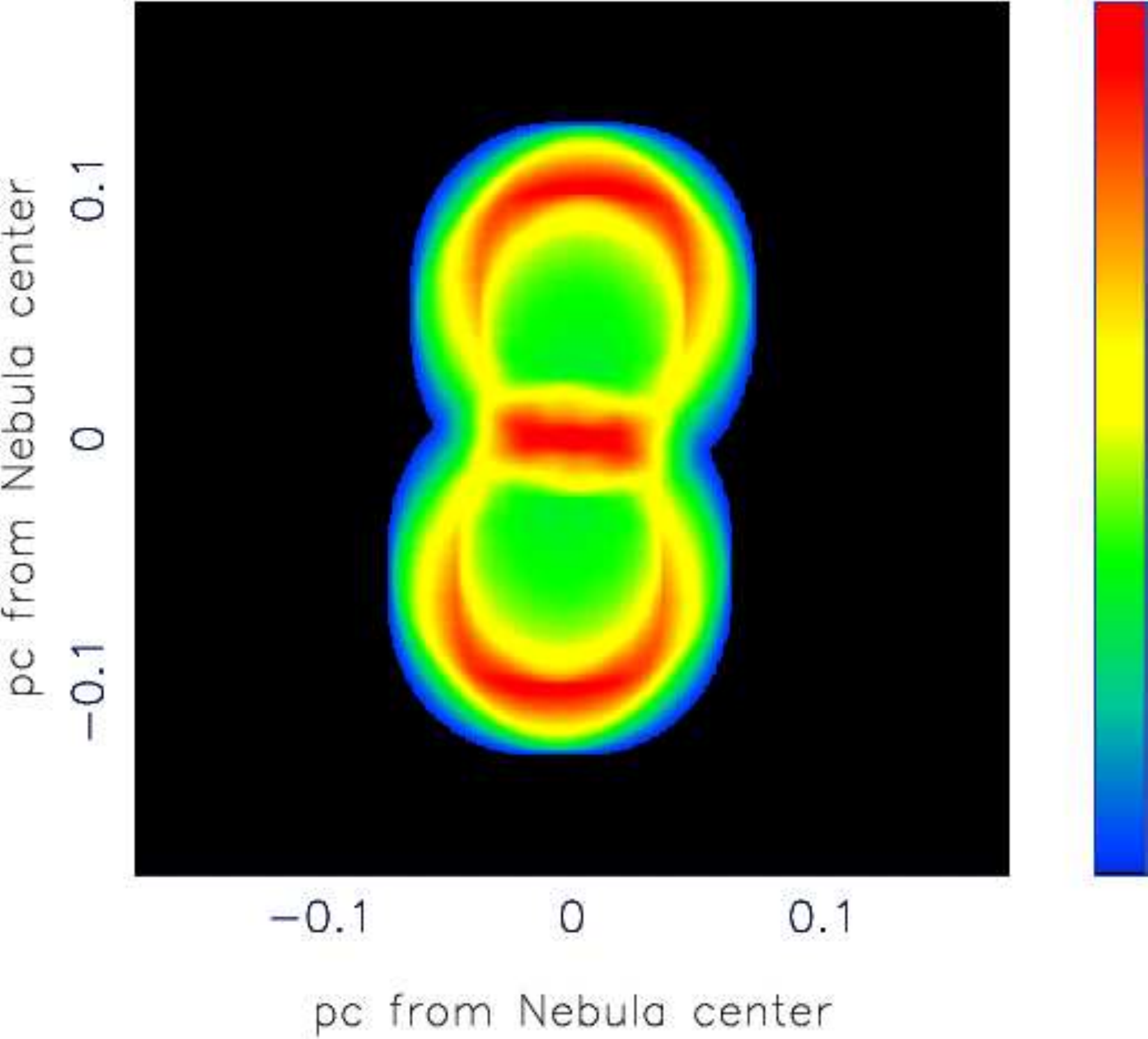}
  \end {center}
\caption {
Map of the theoretical intensity  of the 
rotated MyCn 18 .
Physical parameters as in Table~\ref{parametershom} and
$f$=12 .
The three Eulerian angles 
characterizing the point of view are 
     $ \Phi   $=130     $^{\circ }  $, 
     $ \Theta $=40   $^{\circ }$
and  $ \Psi   $=5     $^{\circ }   $.
          }%
    \label{mycn1840_heat}
    \end{figure}

\begin{figure*}
\begin{center}
\includegraphics[width=8cm]{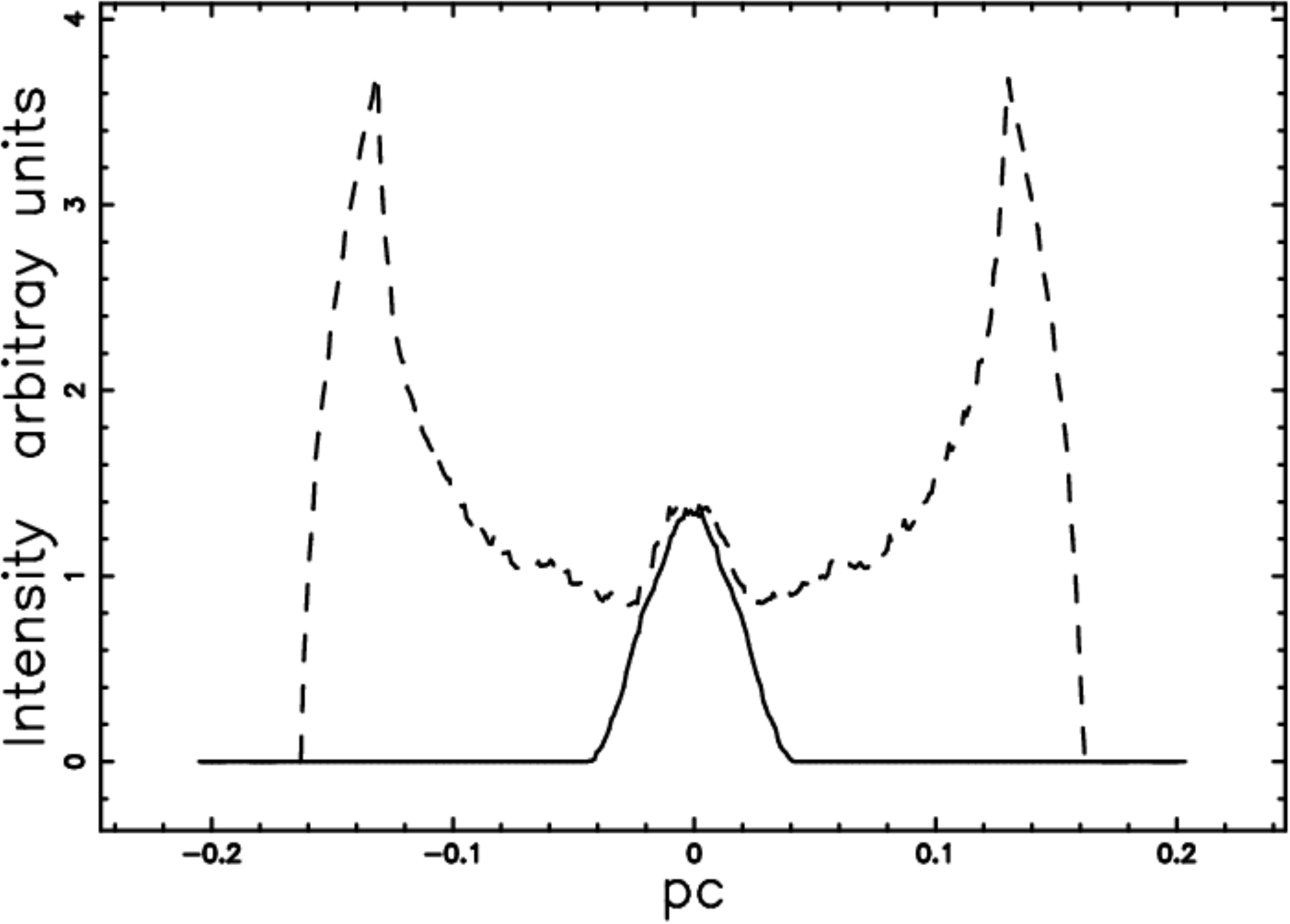}
\end {center}
\caption
{
 Two  cut of the mathematical  intensity ${\it I}$
 crossing the center    of  MyCn 18  :
 equatorial cut          (full line)
 and  polar cut          (dotted line) .
 Parameters as in Figure~\ref{mycn18_heat}.
}
\label{cut_xy_mycn18}
    \end{figure*}

\begin{figure*}
\begin{center}
\includegraphics[width=8cm]{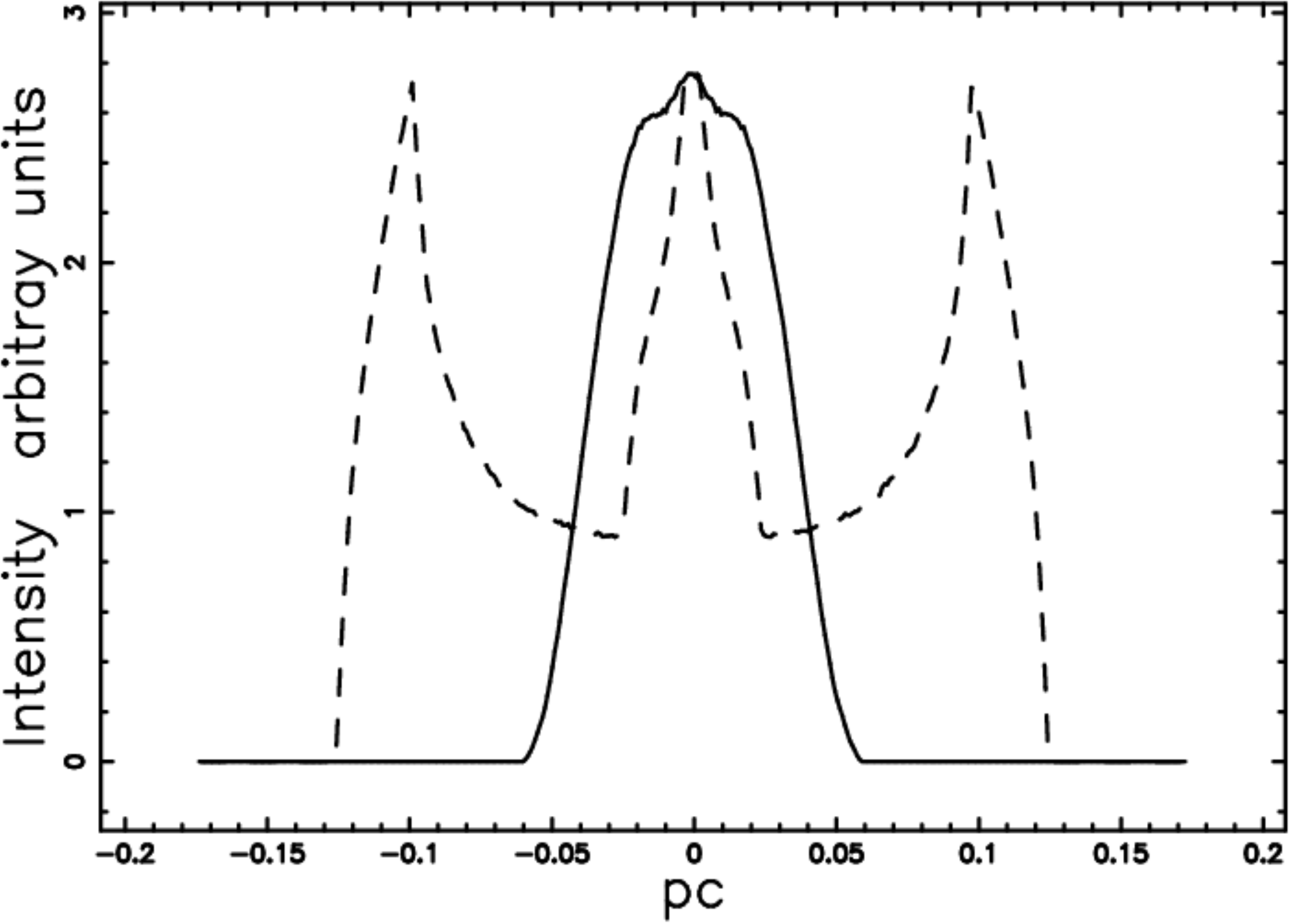}
\end {center}
\caption
{
 Two  cut of the mathematical  intensity ${\it I}$
 crossing the center    of the rotated  MyCn 18 nebula:
 equatorial cut          (full line)
 and  polar cut          (dotted line) .
 Parameters as in Figure~\ref{mycn1840_heat}.
}
\label{cut_xy_mycn1840}
    \end{figure*}

In order to explain some of the morphologies
which characterize the PN's we first map 
MyCn 18    with the polar axis in the vertical 
direction , see map in intensity in 
Figure~\ref{mycn18_heat}. 
The vertical  and horizontal cut in intensity
are reported in Figure~\ref{cut_xy_mycn18}.
The point of view of the observer as modeled 
by the Euler angles increases  the complexity
of the shapes : Figure~\ref{mycn1840_heat}  
reports the after rotation image 
and  Figure~\ref{cut_xy_mycn1840} the vertical and
horizontal rotated cut.
The after rotation image contains the double ring
and an enhancement in  intensity of  the central
region which  characterize
MyCn 18.

 
This central enhancement can be considered
one of the various morphologies that the PNs present
and is similar to  model $BL_1-F$ in Figure~3 
of  the Atlas of synthetic line profiles by \cite{Morisset2008}. 

\section{Conclusions}

{\bf Law of motion}
The law  of motion in the case of a symmetric motion 
can be modeled by the Sedov Solution or the radial momentum
conservation.
These two models allow to determine the approximate 
age of A39   which is 8710 $yr$ for the Sedov solution and 50000 $yr$
for the radial momentum conservation.
In presence of gradients as given  , for example ,
by an exponential behavior, the solution is deduced 
through the radial momentum conservation.
The comparison with the astronomical data is now more complicated
and the single and multiple efficiency  in the radius determination
have  been introduced.
When , for example ,  MyCn 18   is considered ,
the multiple efficiency over 18 directions is $90.66\%$
when the age of 2000 $yr$ is adopted.

{\bf Diffusion} 

The number density in a thick layer 
surrounding the ellipsoid of expansion can be considered 
constant or variable from a maximum value to a minimum
value with the growing or diminishing radius in respect 
to the expansion position.
In the case of a variable number density the framework
of the mathematical diffusion has been adopted,
see formulas~(\ref{cab}) and ~(\ref{cbc}).
The  case of diffusion with drift has been analytically
solved , see formulas~(\ref{cab_drift}) and ~(\ref{cbc_drift}),
and the theoretical formulas have been compared
with values generated by Monte Carlo simulations.
 
{\bf Images}

The intensity of the image of a PN
when the intensity of emission is proportional
to the square of the number density 
can be computed
through
\begin{itemize} 
\item  an analytical evaluation of lines of sight 
when the number density is constant between two spheres
or in one sphere ,
see formula~(\ref{irim})
and  formula~(\ref{isphere}).
\item
analytical evaluation of integrals when the 
number density is variable , see
formulas (~\ref{I_1}), (~\ref{I_2}) and  (~\ref{I_3}),
when the motion  is symmetric.
In this framework it  is also  possible to build a two-phase
diffusion model  that allows us  to reproduce the faint 
extended halo,
see formula~(\ref{isum}).
\item 
a numerical  evaluation of integrals when the 
number density is variable , the drift is present 
 and  the motion  is symmetric, 
see Section~\ref{secdrift}. 
\item
a numerical evaluation of lines of sight  when 
the motion is asymmetric,
see Section~\ref{seccomplex}. 
\end{itemize} 

In the case of A39 the $\chi^2$ of comparison
between the  theoretical and observed  cut in intensity  can
be evaluated , see  Table~\ref{tablechi2}.
From a careful evaluation of Table~\ref{tablechi2}
it is possible to conclude that the models here considered
produce  $\chi^2$  which  are slightly bigger than the rim model
of  \cite{Jacoby2001}.
When , conversely , a diffuse halo is considered 
the  $\chi^2$ is smaller ;  this makes   the study of the interaction
between PN and  the surrounding halo an interesting field  of research.

{\bf Next step}

Here we have 
explored the conservation of the radial momentum 
in a medium with an exponential behavior 
of the type $\rho \propto \exp {- \frac {R\times \sin (\theta) }{h} } $
which  is symmetric in respect to the plane $z=0$ .
The next target can be 
the analysis of the conservation of the radial momentum 
in a spherical symmetry of the type 
$\rho \propto  R^{-\alpha}$ ,
see  Section~\ref{secalfa}.
In this case the  spatially asymmetric motion can be
obtained by considering the 
conservation of the radial momentum  in a medium 
with density of the type 
$\rho \propto R^{-\alpha}\times \exp {- \frac {R\times \sin (\theta) }{h} }
$.

\begin{table}
      \caption{Data of the simulation of the Ring nebula }
         \label{tablechi2}
      \[
         \begin{array}{lcc}
            \hline
            \noalign{\smallskip}
model       &  \mbox{ $\chi^2 $} & \mbox{ $\chi^2 $  }~of~reference,~Jacoby~et~al.~2001 \\ 
rim~with~ fixed~thickness                        &  1.487  & 0.862   \\
diffusion                                        &  19.03  & 12.60   \\
diffusion~with~drift                             &  20.96  & 10.36   \\
diffusion~+~halo                                 &  2.29   & 12.606   \\
            \noalign{\smallskip}
            \hline
         \end{array}
      \]
   \end{table}

\section*{Acknowledgments}

The  astronomical  data
of the intensity profile of the PN A39
were kindly provided by  G. Jacoby .


\label{lastpage}  
\end  {document}